\documentclass[sigconf]{acmart}

\usepackage{multirow}
\usepackage{enumitem}
\usepackage[normalem]{ulem}
\useunder{\uline}{\ul}{}
\usepackage{svg}
\usepackage{subcaption}
\usepackage{threeparttable}

\newtheorem{mydef}{Definition}
\AtBeginDocument{%
  \providecommand\BibTeX{{%
    \normalfont B\kern-0.5em{\scshape i\kern-0.25em b}\kern-0.8em\TeX}}}

\setcopyright{acmcopyright}
\copyrightyear{2023} 
\acmYear{2023} 
\setcopyright{acmlicensed}\acmConference[WSDM '23]{Proceedings of the Sixteenth ACM International Conference on Web Search and Data Mining}{February 27-March 3, 2023}{Singapore, Singapore}
\acmBooktitle{Proceedings of the Sixteenth ACM International Conference on Web Search and Data Mining (WSDM '23), February 27-March 3, 2023, Singapore, Singapore}
\acmPrice{15.00}
\acmDOI{10.1145/3539597.3570406}
\acmISBN{978-1-4503-9407-9/23/02}

\begin{document}
\title{Ranking-based Group Identification via Factorized Attention on Social Tripartite Graph}


\author{Mingdai~Yang}
\affiliation{%
  \institution{University of Illinois at Chicago}
  \city{Chicago}
  \country{USA}}
\email{myang72@uic.edu}

\author{Zhiwei~Liu}
\affiliation{%
  \institution{Salesforce AI Research}
  \city{Palo Alto}
  \country{USA}
}
\email{zhiweiliu@salesforce.com}

\author{Liangwei~Yang, Xiaolong~Liu, Chen~Wang}
\affiliation{%
  \institution{University of Illinois at Chicago}
  \city{Chicago}
  \country{USA}}
\email{{lyang84, xliu262, cwang266}@uic.edu}

\author{Hao Peng}
\affiliation{%
   \institution{School of Cyber Science and Technology, Beihang University,}
   \country{Beijing, China}}
\email{penghao@buaa.edu.cn}
\authornote{Corresponding author}

\author{Philip S.~Yu}
\affiliation{%
  \institution{University of Illinois at Chicago}
  \city{Chicago}
  \country{USA}}
\email{psyu@uic.edu}

\renewcommand{\shortauthors}{Trovato and Tobin, et al.}

\begin{abstract}
Due to the proliferation of social media, a growing number of users search for and join group activities in their daily life. This develops a need for the study on the ranking-based group identification (RGI) task, i.e., recommending groups to users. 
The major challenge in this task is how to effectively and efficiently leverage both the item interaction and group participation of users' online behaviors.
Though recent developments of Graph Neural Networks (GNNs) succeed in simultaneously aggregating both social and user-item interaction, they however fail to comprehensively resolve this RGI task. 
In this paper, we propose a novel GNN-based framework named \textbf{C}ontextualized \textbf{F}actorized \textbf{A}ttention for \textbf{G}roup identification (CFAG). 
We devise tripartite graph convolution layers to aggregate information from different types of neighborhoods among users, groups, and items. 
To cope with the data sparsity issue, we devise a novel propagation augmentation~(PA) layer, which is based on our proposed factorized attention mechanism. 
PA layers efficiently learn the relatedness of non-neighbor nodes to improve the information propagation to users.
Experimental results on three benchmark datasets verify the superiority of CFAG. 
Additional detailed investigations are conducted to demonstrate the effectiveness of the proposed framework.
\end{abstract}





\maketitle

\section{Introduction}

\begin{figure}[t]
  \centering
  \includegraphics[width=\linewidth]{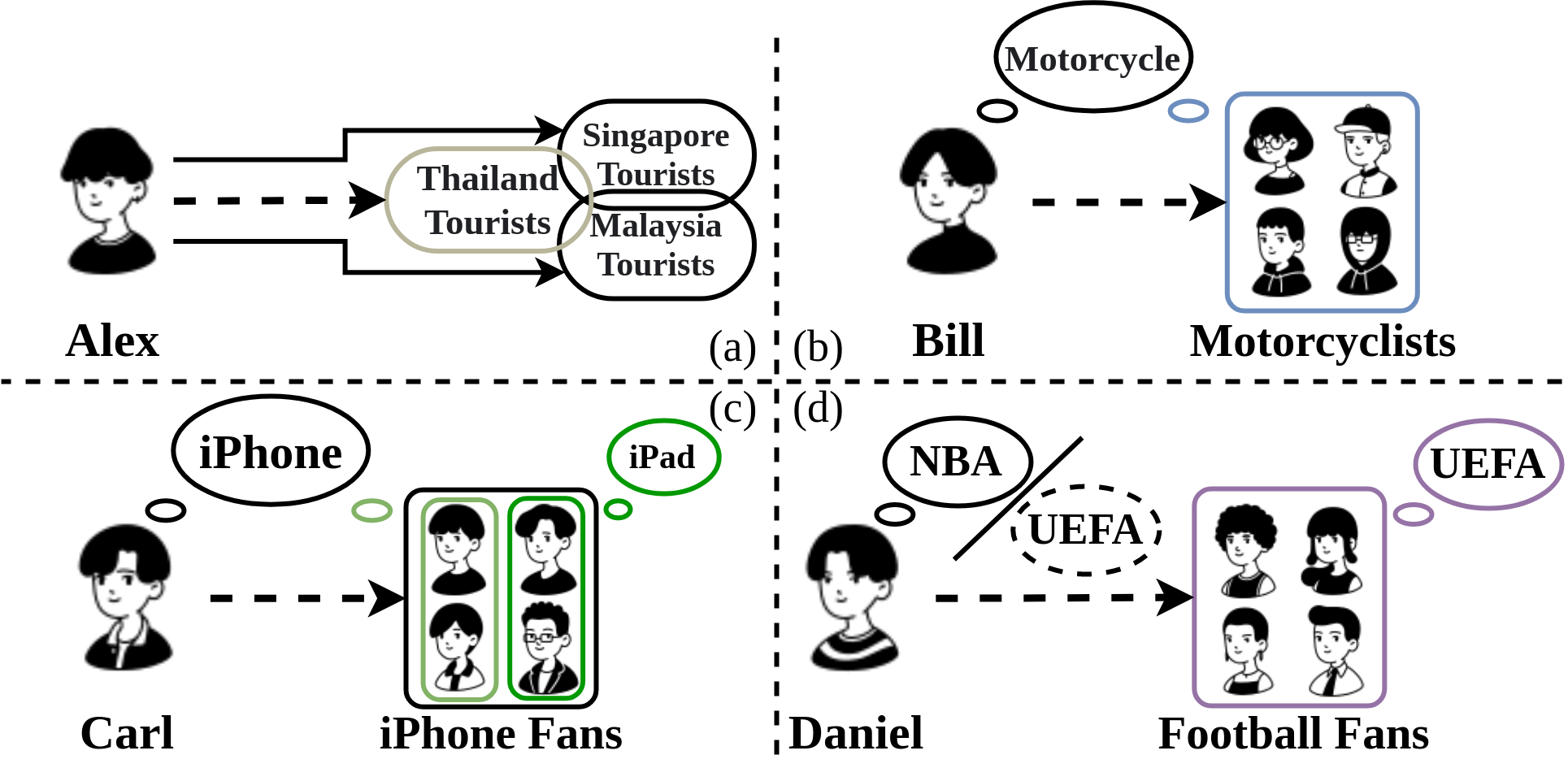}
  \caption{Illustration of group identification.}
  \label{illustrative example}
\end{figure}

Given the difficulties in information collection and decision-making among the overwhelming options for individuals, a growing number of customers prefer to join specific groups for suggestions in advance of consumption.
For example, on the Steam video game platform\footnote{\url{https://store.steampowered.com/}}, if a video game player hesitates about whether she should purchase a newly released game, she would seek suggestions from a group of players who have played that game for a period.
Online groups offer spaces for users to share experiences, which in turn provides a reference to other group members. It assists them in locating their demands accurately and even influences their interests in items. 
We illustrate a toy example in Figure~\ref{illustrative example}, Alex is interested in Motorcycle. Hence, he would like to join a Motorcyclists group.
In terms of platforms, users' attachment to online groups can significantly increase the participation and retention rate of their users~\cite{engagement@17}. 
Compared to promoting content directly to potential users ~\cite{promotion@12}, recommending users to groups based on their interests is a more feasible way to help platforms build emotional bonds with users for maintaining long-term stickiness, which is however under-explored.

Most existing recommender systems have been widely adopted to discover relevant content~\cite{park2017deep}, products~\cite{hao2020p} or services~\cite{ISINKAYE@15}.
The recent successes of GNNs~\cite{gat17,hamilton2017inductive} inspire graph-based recommender systems~\cite{lightgcn20,ngcf19}.
These graph-based recommender systems focus on bipartite recommendation tasks, e.g., user-item recommendation~\cite{ngcf19,lightgcn20,sgl21}.
These methods allow information to propagate over graph through high-order connectivity. 

However, directly applying those bipartite graph-based approaches to user-group interactions for prediction is far from comprehensive. 
In this paper, we argue that successfully recommending groups to users requires combined modeling of both group participation and item interactions.
On one hand, users may be interested in a group similar to groups they joined before. 
We demonstrate this via a toy example in Figure~\ref{illustrative example}(a). Alex visited Singapore and Malaysia before and became a member of corresponding tourist groups. Therefore, he would like to join a tourist group in Thailand as all of them are in Southeast Asia. 
On the other hand, users may join a new group because of item interests, even though never participating in any relevant groups before. 
For example in Figure~\ref{illustrative example}, Bill is fond of motorcycles. As such, we should recommend him a motorcycle riders group as Figure~\ref{illustrative example}(b) demonstrated.
Therefore, only when simultaneously characterizing both group participation and item interactions can we develop a satisfying group identification system. 





In this work, we define recommending groups to users as ranking-based group identification (RGI) problem. 
RGI problem is distinguishable from group recommendation~\cite{groupim@20,agree18,yuan14} and community detection~\cite{fortunato2010community,Jin19}, where
the former is to recommend items to groups and/or users by leveraging group information while the latter is to cluster users.
As aforementioned, incorporating both item interactions and group participation is necessary for RGI problem, which is rather challenging from two perspectives.

Firstly, we should consider high-order connectivity in RGI problem. 
We demonstrate the high-order connectivity in Figure~\ref{illustrative example}(c), where Carl is considering buying a newly-released iPhone. 
Thus, he joins the \textit{iPhone Fans} group to discuss with others. 
In this group, other members also have interactions with \textit{iPad}.  
As such, Carl may also have potential interests in groups relevant to iPad. 
Though Carl has no direct connection with iPad, we can still discover this potential interest through high-order connections, \textit{i.e.}, user-group-user-item-group in this case. 
Existing works on recommender systems harness additional preferences (\textit{i.e.} item interactions in this paper) as side information~\cite{agree18,han@18,graphRec}, which is unable to resolve high-order connectivity.

Secondly, the relatedness of items is crucial for identifying users' group interests. To be more specific, a user may be interested in a group if its members share similar items' interests with this user, even though not exactly the same item that this user has interactions with. 
We illustrate this in Figure~\ref{illustrative example}(d).
Daniel subscribes to NBA basketball games stream\footnote{\url{https://www.espn.com/nba/}}. 
He would be interested in a Football Fans group where the members are all football fans having UEFA football games stream\footnote{\url{https://www.uefa.com/uefaeuropaleague/}} subscriptions as a result of the relatedness between NBA and UEFA.

To this end, we investigate the RGI problem upon a social tripartite graph, which includes
user, group, and item as nodes and their associated interactions as edges.
We decide to propagate information over this graph to learn representations of nodes. 
The designing reasons are threefold. First, this tripartite graph incorporates both group participation and
item interactions. Second, propagating information over a social tripartite graph enables the high-order connectivity characterization. Third, similar representations are able to reveal the relatedness of RGI problem.

However, directly adopting existing methods for RGI problem is not suitable since we observe severe sparsity issues. 
In other words, many users have few group/item interactions. 
Under this circumstance, aggregation over the tripartite graph is less likely to learn high-quality representations of nodes due to the unavailability of adequate propagation paths. 

To this end, we propose a novel framework, named Contextualized Factorized Attention for Group
identification (CFAG). 
CFAG is a GNN over our proposed social tripartite graph. To resolve the sparsity issue, we propose to endow the model with the ability of discovering more potential interactions of users during training via factorized attention mechanism. 
More concretely, we propose learning weights between users and \textit{all} other groups/items that they have no interaction with. 
In this way, the information from potential groups and items is able to propagate to users. 
The weights reflect the potential impacts of those groups/items and construct the propagation paths. 
Nonetheless, training all the pair-wise weights is of high time- and space- complexity. 
Instead, we learn contextual embeddings for both groups and items. Then, the weights between a user and all candidate groups and items are inferred based on the embedding similarity between candidates and users' participated groups and interacted items, respectively. 
Finally, the weights between this user and candidates are calculated as the attention score.
To train a CFAG, besides the contextual embedding for items and groups, all nodes have personalized embeddings.
We iteratively update the personalized embedding and contextual embedding via propagation over the graph.
The rankings score between a user and a group is based on their final personalized embeddings from the output layer. 
And we optimize the framework based on BPR loss~\cite{MF09}.
We highlight our key contributions as follows:
\begin{itemize}[leftmargin=*]
    \item We propose a novel framework CFAG, a GNN-based model for group identification, which can propagate the information on the social tripartite graph and conduct recommendation.
    \item We devise novel propagation augmentation layers with factorized attention mechanism in CFAG to cope with the sparsity issue, which explores non-existing interactions and enhances the propagation ability on graphs with high sparsity.
    \item We collect and release one large dataset for RGI task. We conduct extensive experiments on this dataset and two other public datasets. 
    The significant improvements of CFAG on all datasets indicate its superiority in tackling the RGI problem.
\end{itemize}

\section{Related Works}
Since there is no previous work targeting the exact same task, i.e., recommending groups to users based on  graphs, we will introduce some closely related work: (1) Community detection, which shares similar goals with our task, is introduced in Section~\ref{Community Detection}. (2) Recommender systems that utilize users' social information are reviewed in Section~\ref{RS Incorporating Social Information}. (3) With similar deep learning techniques, GNN-based methods are introduced in Section~\ref{Graph Neural Network-Based RS}. 
In each subsection, we discuss their relationships to the proposed RGI task.

\subsection{Group Recommender Systems}\label{Community Detection}
Group recommender systems refer to recommending groups to their potential members. 
Traditional group recommender systems apply various algorithms to recover user-group membership matrices with available side information.
For example, semantic information from descriptions of groups~\cite{chen08} and visual information from photos shared by users~\cite{wang12} can be incorporated with a collaborative filtering framework to perform personalized group recommendations.
User behaviors in different time periods~\cite{wang16, yang21}, such as joining groups, can also be leveraged for recommending groups to users. 
However, the requirement of side information degrades the performance of those methods when recommending groups to users with only interaction information. 
Some recent works~\cite{grouprec21, Telegram21} investigate recommending groups to users with only user-group interactions. They directly characterize the bipartite structure between users and groups, while item interaction information is ignored.

Besides recommending groups to users, the term \textit{group recommendation} in literature also refers to recommending items to a group of users~\cite{groupim@20, agree18}, which differs from the focus in this paper.

\subsection{Social Information-based RS} \label{RS Incorporating Social Information}
Social relations between users have been applied to the recommender system to alleviate data sparsity problem~\cite{ConsisRec,Hu19}. Research in social recommendation combines a user-user graph and the bipartite user-item graph to better understand users’ preferences on items~\cite{sr_1,social_3}. Friend recommendation is another research topic based on the homogeneous graph formed by social relations to find possible social links between users~\cite{fr_1,fr_2}. For these RS involving social relations among users, the main challenge lies in how to model the influence between users~\cite{GLRSreview}. Typical approaches addressing this challenge contain random walk~\cite{sr_rw1}, GAT~\cite{graphRec} and graph embedding~\cite{sr_emb}. 

In contrast to using social information to predict possible links among user-user pairs or user-item pairs, our work focuses on predicting links between users and groups. Different from user-user or user-item relation, user-group relation could be entangled promiscuously~\cite{yuan14,agree18}.

\subsection{Graph-Based RS}\label{Graph Neural Network-Based RS}
GNN has been widely leveraged to address the most important challenges in RS nowadays given its powerful capability of learning informative representations in graph data. The GNN-based methods broadly fall into three classes from the model perspective: (i) Graph Convolutional Network based RS~\cite{ngcf19,lightgcn20,sgl21}; (ii) Graph Attention Network based RS~\cite{gat17,mgat@20}; and (iii) Gated Graph Neural Network based RS~\cite{grurs_3}.

Most previous works on Graph-based RS only focus on user-item bipartite graph~\cite{lightgcn20}. It is difficult to directly apply these works to group identification tasks with three relations to be managed: user-item, group-item and user-group relations. 

Some graph-based works in group-item recommendation also model different interactions between users, items and groups~\cite{agree18}. However, the key challenge of those works is how to aggregate the preference of group members~\cite{agree18,groupim@20}. Our task differs significantly in that our goal is to predict the preference of individual users on groups he/she never interacted with.

\section{Preliminaries}

\begin{figure*}[h]
  \centering
    \includegraphics[width=0.99\linewidth]{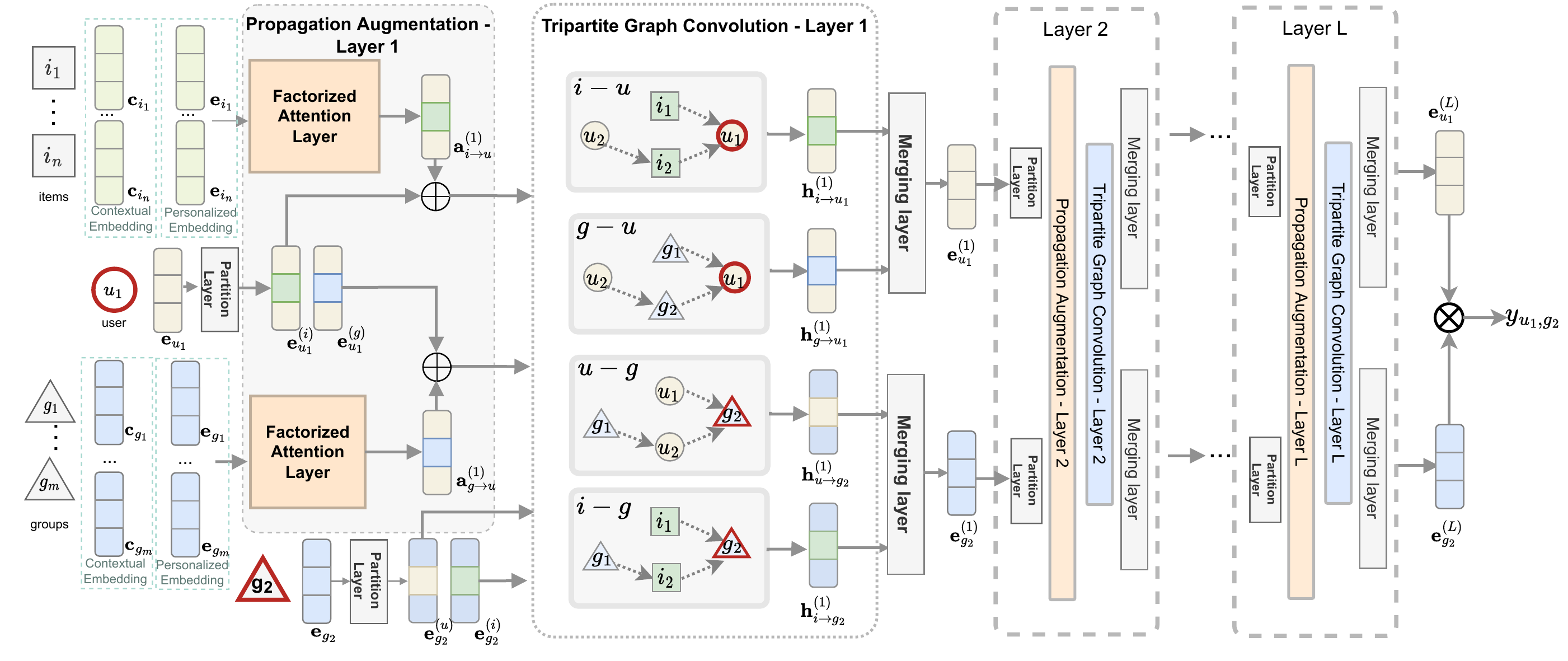}
  \caption{The framework of CFAG. This is the computation process when inferring the score between user $u_1$ and group $g_2$. }
  \label{fig:framework}
\end{figure*}

In this section, we first define the social tripartite graph and then formulate the problem of rank-based group identification (RGI). 
\begin{mydef}
\textbf{(Social Tripartite Graph).} Given three disjoint nodes, i.e, user nodes set $\mathcal{U}$, group nodes set $\mathcal{G}$, and item nodes set $\mathcal{I}$, and given their interactive edges, i.e. user-group edges $E_{\mathcal{U},\mathcal{G}}$, user-item edges $E_{\mathcal{U},\mathcal{I}}$, and group-item edges $E_{\mathcal{G},\mathcal{I}}$, we define a social tripartite graph as $\mathcal{T}=\{V,E\}$, where $V = \mathcal{U} \cup \mathcal{I} 
\cup \mathcal{G}$ and $E = E_{\mathcal{U},\mathcal{G}} \cup E_{\mathcal{U},\mathcal{I}} \cup E_{\mathcal{G},\mathcal{I}}$.
\end{mydef}
The social tripartite graph is an undirected and unweighted heterogeneous graph. In the following sections, we denote user nodes as $u$, group nodes as $g$, and item nodes as $i$. An edge $(u_j,g_k) \in E_{\mathcal{U},\mathcal{G}}$ in the graph represents that user $u_j\in \mathcal{U}$ is a member of a group $g_k\in \mathcal{G}$. Similarly, an edge in $E_{\mathcal{U},\mathcal{I}}$ or $E_{\mathcal{G},\mathcal{I}}$ represents that a user or a group shows preference on an item.
Note that if we have no item set available, \textit{i.e.} $\mathcal{I} = \emptyset$, this tripartite graph degenerates to a bipartite graph, which is utilized in existing recommender systems. 

During computation, we use adjacency matrices to represent the edges. Specifically, we define the user-group adjacency matrix $\pmb{X} = [x_{jk}]\in\mathbb{R} ^{|\mathcal{U}|\times |\mathcal{G}|}$ to represent the user-group interactions, where $x_{jk} = 1$ if an edge $(u_j,g_k)$ exists in the social tripartite graph, i.e., user $u_j$ is a member of group $g_k$, and otherwise $x_{jk} = 0$. 
Analogously,  we define user-item and group-item adjacency matrices as $\pmb{Y}\in\mathbb{R}^{|\mathcal{U}|\times |\mathcal{I}|}$ and $\pmb{Z} \in\mathbb{R}^{|\mathcal{G}|\times |\mathcal{I}|}$, respectively. 
Then, the target is to conduct ranking-based edge prediction, which is defined as:
\begin{mydef}
\textbf{(Ranking-based Group Identification)}. Given a social tripartite graph $\mathcal{T}$, the ranking-based group identification (RGI) for a user $u$ is to predict a ranking list of groups $\{g_1, g_2, \dots, g_k\}$, with which this user has no interactions.
\end{mydef}
In other words, we recommend a list of groups that this user $u$ is of potential interest in RGI. Note that we distinguish the group as another entity rather than a simple union of users due to its special characteristics, e.g. group information. 



\section{Method}
In this section, we present the proposed CFAG model for the group identification task. The framework of CFAG is shown in Figure~\ref{fig:framework}. We start by introducing all embedding layers to be trained in this framework.
Next, we demonstrate how to aggregate the interactions from different neighborhoods on the social tripartite graph. 
Specifically, we adopt tripartite graph convolution networks to learn the user and group personalized embeddings for recommending groups. 
We further propose a Factorized Attention module for user-item interactions and user-group interactions to infer the relevance of each group and each item to the target user.
\subsection{Embedding layer}
We maintain an embedding layer $\mathbf{E}\in \mathbb{R}^{d\times(|\mathcal{U}|+|\mathcal{G}|+|\mathcal{I}|)}$, where each column represents the trainable personalized embedding for each node $v\in V$ in the graph. 
In addition to the personalized embedding, we also have contextual embedding layers for both items and group, denoted as $\mathbf{C}_{i}\in\mathbb{R}^{d\times |\mathcal{I}|}$ and $\mathbf{C}_{g}\in\mathbb{R}^{d\times |\mathcal{G}|}$, respectively. They are used and trained for Factorized Attention to infer the influence between groups (items) for each user, respectively. More details will be presented in Sec.~\ref{sec:FacAtt}.

\subsection{Tripartite Graph Convolution}\label{TriConv}

GNNs learn node embeddings by propagating information from neighbors to center nodes. 
However, existing GNN layers are not suitable due to the heterogeneity of the social tripartite graph, i.e. the distinctions of different types of nodes. 
Hence, we devise a novel tripartite graph convolution  to propagate information between users, groups and items. 
Instead of directly aggregating embeddings of neighbors,  we employ \textit{partition layers} to divide neighbor information as two branches for aggregation.
In the following part of this section, we explain each module of tripartite graph convolution over the center node group $g$.
The tripartite graph convolution over user and items can be derived in analogy. 
To be more specific, partition layers divide group information $\mathbf{e}_{g}$ as group-user information $\mathbf{e}_{g}^{(u)}$ and group-item information $\mathbf{e}_{g}^{(i)}$ for the following propagation to users and items, respectively:
\begin{equation}\label{eq:partition_layer}
    \mathbf{e}_{g}^{(u)}, \mathbf{e}_{g}^{(i)} = \textbf{PT}(\mathbf{e}_{g}),
\end{equation}
where \textbf{PT}$()$ denotes the partition layer.
We justify various partition layers, for example \textbf{PT}$()$ can be two linear transformations, \textit{i.e.}, $\mathbf{e}_{g}^{(u)}=\mathbf{W}^{(u)}_{g}\mathbf{e}_{g}$,  $\mathbf{e}_{g}^{(i)}=\mathbf{W}^{(i)}_{g}\mathbf{e}_{g}$. 
Moreover, it can be a simple division over the dimension, \textit{i.e.}, $[\mathbf{e}_{g}^{(u)} \| \mathbf{e}_{g}^{(i)}] = \mathbf{e}_{g}$, where $\|$ is the concatenation of two embeddings.  
In fact, experiments in Sec.~\ref{sec:ablation} justify the superiority of later simple division over the dimension.
The partition layers distinguish the impacts of different types of neighbors. 
Examples of dividing the information of group $g_2$ and user $u_1$ via partition layers are illustrated in Figure~\ref{fig:framework}. 

Hereafter, we learn node embeddings by aggregating corresponding neighboring information from partition layers. 
The group information aggregated from both neighbor users and items are as follows:
\begin{equation}\label{eq:information_propagation}
\begin{aligned}
\mathbf{h}_{u\rightarrow g} &= \textbf{AGG}(\{\mathbf{e}_u^{(g)}|u\in\mathcal{N}_g^{(u)}\}) \\
\mathbf{h}_{i\rightarrow g} &= \textbf{AGG}(\{\mathbf{e}_i^{(g)}|i\in\mathcal{N}_g^{(i)}\}),
\end{aligned}
\end{equation}
where $\mathbf{h}_{u\rightarrow g}$ and $\mathbf{h}_{i\rightarrow g}$ denote the user and item information propagated to group $g$, respectively.
The \textbf{AGG}() represents the aggregation layers, such as the GCN~\cite{GCN16} aggregation. 
These aggregation layers propagate associated information to group $g$. 
For example, user-to-group information $\mathbf{h}_{u\rightarrow g}$ is aggregated from the group partition of user embeddings, denoted as $\mathbf{e}_{u}^{(g)}$. 
$\mathcal{N}_{g}^{(u)}$ represents all the neighbor users of group $g$. 
In analogy, item-to-group information $\mathbf{h}_{i\rightarrow g}$ is aggregated from the group partition of neighbor item embeddings, \textit{i.e.}, $\mathbf{e}_{i}^{(g)}$. 

Next, we combine the information from two branches as one embedding for group $g$ via \textit{merging layers} \textbf{MG}$()$ as follows:
\begin{equation}\label{eq:merging_layer}
    \mathbf{e}_{g}^{*} = \textbf{MG}(\mathbf{h}_{u\rightarrow g},\: \mathbf{h}_{i\rightarrow g}),
\end{equation}
where $\mathbf{e}_{g}^{*}$ is the output embedding for group $g$.  We demonstrate the merging layers in Figure~\ref{fig:framework}.
In this paper, we investigate different types of \textbf{MG}$(\cdot,\: \cdot)$ layers, including direct concatenation, concatenation before fully-connection, and concatenation after fully-connection.

\textbf{Multi-layer propagation.} By stacking multiple those layers in Eq.~(\ref{eq:partition_layer}), (\ref{eq:information_propagation}) and (\ref{eq:merging_layer}), we construct the multi-layer propagation pattern of tripartite graph convolution and rewrite the Eq.~(\ref{eq:merging_layer}) as follows:
\begin{equation}\label{eq:multi-layer}
    \mathbf{e}_{g}^{(l)} = \textbf{MG}(\mathbf{h}_{u\rightarrow g}^{(l)},\: \mathbf{h}_{i\rightarrow g}^{(l)}),
\end{equation}
where $\mathbf{h}_{u\rightarrow g}^{(l)}$ and $\mathbf{h}_{i\rightarrow g}^{(l)}$ denote the information propagation in Eq.~(\ref{eq:information_propagation}) on the $l$-th layer.
In Figure~\ref{fig:framework}, we stack $L$ layers and yield the final embedding $\mathbf{e}_{g_2}^{(L)}$ and $\mathbf{e}_{u_1}^{(L)}$ for group $g_2$ and user $u_1$, respectively. 

\subsection{Propagation Augmentation}\label{sec:FacAtt}
\begin{figure}
  \centering
  \includegraphics[width=0.7\linewidth]{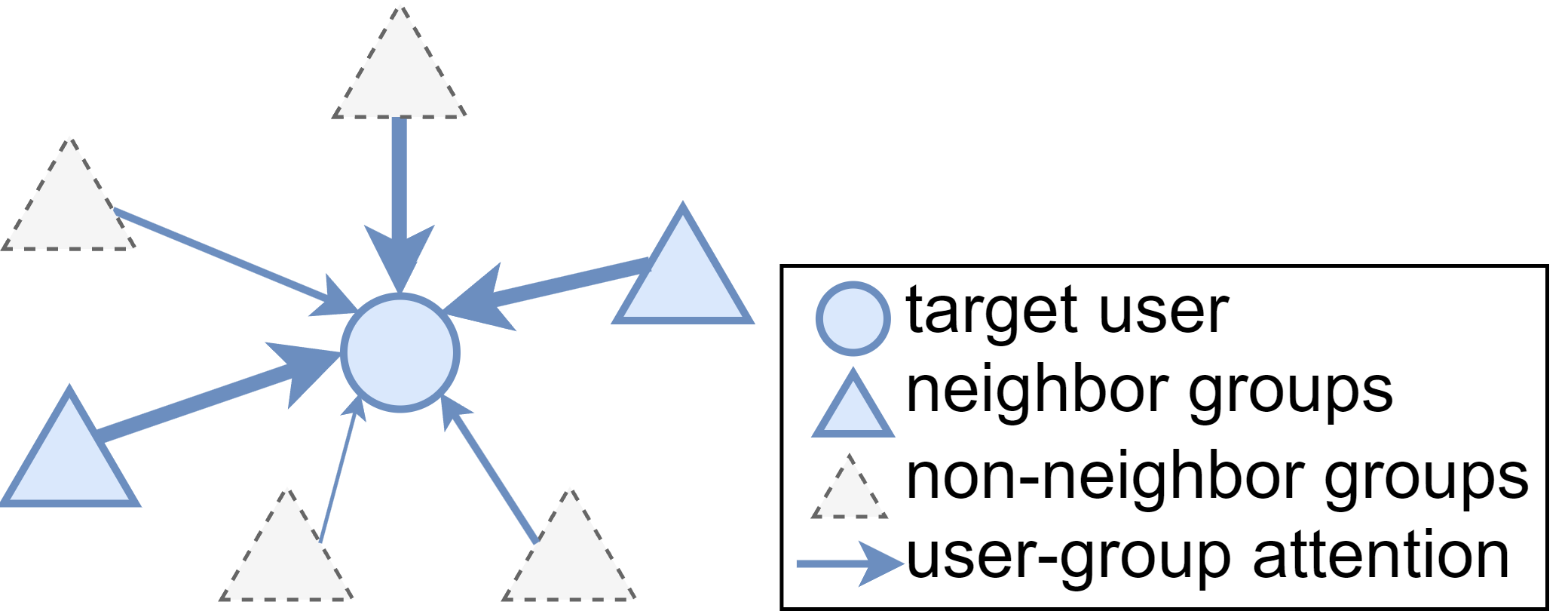}
  \caption{Illustration of propagation augmentation on group-user neighbors. We aggregate all groups via learned group contextualization attention.}
  \label{fig:PA}
\end{figure}

In RGI problem, we observe severe sparsity issues in the graph, \textit{e.g.} few group participation for users.
Those sparsity issues impair the propagation over the graph, and thus, those nodes with few neighbors are unable to be well trained. 
To this end, we propose a propagation augmentation~(PA) layer before the tripartite graph convolution. 
Intuitively, PA layers aggregate the information from all \textit{non-neighbor} nodes to the center node.
As such, it augments the propagation paths for sparse graphs. 
An illustration of the PA layer on group-to-user propagation is in Figure~\ref{fig:PA}. 
The dash triangles represent those unconnected groups of the target user. 
We assign them the user-group attention weights as in the solid lines.
PA layer constructs additional context for the target user by exploring non-neighbor groups.
Hence, we have additional user-to-group information as follows:
\begin{equation}\label{eq:user-to-group info augmentation}
\mathbf{a}_{g\rightarrow u} =
\sum\nolimits_{g \in \mathcal{G}}\alpha_{gu}
\mathbf{e}_{g}^{(u)},
\end{equation}
where $\mathbf{a}_{g\rightarrow u}$ denotes the addition information propagated to the target user from all groups. The $\alpha_{gu}$ is the attention weight from group $g$ to user $u$. And $\mathbf{e}_{g}^{(u)}$ represents the user partition for group personalized embedding $\mathbf{e}_{g}$. 
Analogously, we also have additional user-to-item information as follows:
\begin{equation}
\mathbf{a}_{i\rightarrow u} =
\sum\nolimits_{i \in \mathcal{I}}\alpha_{iu}
\mathbf{e}_{i}^{(u)},
\end{equation}
where $\alpha_{iu}$ is the attention weight from item $i$ to user $u$. And $\mathbf{e}_{i}^{(u)}$ represents the user partition for item personalized embedding $\mathbf{e}_{i}$. 

However, it is problematic if we directly learn these pair-wise attention weights. 
For example, the attention weights between users and groups $\alpha_{gu}$ increase the parameter complexity by $\mathcal{O}(|\mathcal{G}|\times |\mathcal{U}|)$, which is rather large and unable to scale. 
Inspired by matrix factorization~\cite{MF09}, we propose a novel \textbf{factorized attention} mechanism to learn those attention weights. 
Basically, groups and items have additional \textit{contextual embeddings} as $\mathbf{C}_{g}$ and $\mathbf{C}_{i}$, respectively, which are utilized to infer corresponding attention weights.
Next, we will introduce the factorized attention mechanism. 
For simplicity, we only explain how to infer group-user attention weight $\alpha_{gu}$ via context embedding of groups. 
Item-user attention weight $\alpha_{iu}$ can be derived in analogy. 

We calculate the attention weight $\alpha_{gu}$ based on the relatedness between this group $g$ and all the neighboring groups of the target user $u$.
The intuition is that users tend to be more interested in groups similar to users' previously participated groups.
We formulate the calculation as follows:

\begin{equation}
\alpha_{gu} = \frac{\text{exp}(\sigma(\sum_{m\in\mathcal{N}^{(g)}_{u}}\mathbf{R}_{mg}))}
{\sum_{k \in \mathcal{G}}\text{exp}(\sigma(
\sum_{m\in\mathcal{N}^{(g)}_{u}}\mathbf{R}_{mk}))},
\end{equation}
where $\mathbf{R}_{mg}$ denotes the relatedness between group $m$ and group $g$ and group $m$ is from user participated groups $\mathcal{N}^{(g)}_{u}$. We use LeakyRelu as the nonlinear function $\sigma(\cdot)$.
The group contextual attention matrix $\mathbf{R}\in \mathbb{R}^{|\mathcal{G}|\times|\mathcal{G}|}$ enables us to aggregate non-direct group information thus augmenting the propagation.
We calculate elements $\mathbf{R}_{mg}\in \mathbf{R}$ as follows:
\begin{equation}\label{eq:relatedness_matrix}
\mathbf{R}_{mg} = \frac{\text{exp}(\mathbf{c}_{m}\cdot\mathbf{c}_{g})}
{\sum_{k\in\mathcal{G}}\text{exp}(\mathbf{c}_{k}\cdot\mathbf{c}_{g})},
\end{equation}
where $\mathbf{c}_{m}\in\mathbb{R}^{d}$ is the contextual embedding for group $m$.
In fact, we calculate the relatedness matrix directly through the product between the group contextual embeddings and its transpose as follows:
\begin{equation}\label{eq:factorization}
\mathbf{R} = \text{softmax}(\mathbf{C}_{g}^\top\mathbf{C}_{g}),
\end{equation}
where $\mathbf{C}_{g}$ is the contextual embeddings for groups. 
As observed, we decompose the calculation of pair-wise attention weights, \textit{i.e.} the $\alpha_{gu}$ in Eq.~(\ref{eq:user-to-group info augmentation}), into the product of a $d$-rank matrix and its transpose.
Hence, it is a factorized attention mechanism. 
Our experiments also demonstrate its better performance compared with existing graph attention layers~\cite{gat17}.
We present the factorized attention layer in Figure~\ref{fig:framework}.
By stacking PA layer with the tripartite convolution layer, we enhance the group and item partition embedding of user $u$ with this additional information propagation as follows: 
\begin{equation}
    \hat{\mathbf{e}}_{u}^{(g)} = \mathbf{e}_{u}^{(g)}+\beta\mathbf{a}_{g\rightarrow u}, \:  \hat{\mathbf{e}}_{u}^{(i)} = \mathbf{e}_{u}^{(i)}+\beta\mathbf{a}_{i\rightarrow u},
\end{equation}
where $\beta\in[0,1]$ is a scalar hyper-parameter to control the intensity of PA. 
Note that when $\beta=0$, there is no augmentation for the propagation. 
Then, we aggregate these enhanced embeddings in tripartite graph convolution and substitute the $\mathbf{e}_{u}^{(g)}$ and $\mathbf{e}_{u}^{(i)}$ in  Eq.~(\ref{eq:information_propagation}) with $ \hat{\mathbf{e}}_{u}^{(g)}$ and $\hat{\mathbf{e}}_{u}^{(i)}$, respectively.
We present this PA layer process in Figure~\ref{fig:framework}.

\subsection{Ranking Prediction and Optimization}\label{RanPre}
The final prediction process is demonstrated in the right part of Figure~\ref{fig:framework}. After stacking $L$ layers, we obtain the user embedding $\mathbf{e}_u^{(L)}$ and group embedding $\mathbf{e}_g^{(L)}$. The ranking score of the user-group pair $(u,g)$ is calculated by the inner product:
\begin{equation}
{y}_{u,g} = \mathbf{e}_u^{(L)}\cdot \mathbf{e}_g^{(L)}.
\end{equation}
Then we employ the pairwise Bayesian Personalized Ranking (BPR) loss~\cite{MF09} as our loss function:
\begin{equation}
\mathcal{L}=\sum\limits_{(u,g,g')\in \mathcal{D}} -\log\sigma(\hat{y}_{u,g}-\hat{y}_{u,g'}) + \lambda \|\Theta\|^2_2,
\end{equation}
where $\mathcal{D}=\{(u,g,g')|g\in \mathcal{G}^{+}_{u}, g'\in \mathcal{G}\backslash\mathcal{G}^{+}_{u}\}$ is the training data with positive interactions and random negative samples.
$\Theta$ is all trainable parameters in the framework, which is regularized by $\lambda$. Adam~\cite{adam14} is chosen as the optimizer.
\section{Experiment}

\subsection{Experimental Setup}

\begin{table*}[!ht]
\caption{RGI performance comparison on three datasets.}\label{tab:performance}
\begin{tabular}{l|cccc|cccc|cccc}
\hline
Dataset     & \multicolumn{4}{c|}{Mafengwo}        & \multicolumn{4}{c|}{Weeplaces}                            & \multicolumn{4}{c}{Steam}                      \\ \hline
Metric      & R@10    & R@20    & N@10   & N@20    & R@10         & R@20         & N@10         & N@20         & R@10    & R@20         & N@10   & N@20         \\ \hline
AGREE       & 0.0287  & 0.0495  & 0.0104 & 0.0157  & 0.0174       & 0.0322       & 0.0081       & 0.0119       & 0.0497  & 0.0683       & 0.0239 & 0.0283       \\
MF-BPR      & 0.1930  & 0.2388  & 0.1185 & 0.1305  & 0.2349       & 0.2826       & 0.1479       & 0.1625       & 0.2034  & 0.2859       & 0.1138 & 0.1356       \\
ENMF &
  {\ul 0.3038} &
  {\ul 0.3568} &
  {\ul 0.1805} &
  {\ul 0.1959} &
  0.2283 &
  0.3260 &
  0.1258 &
  0.1577 &
  {\ul 0.2424} &
  0.3153 &
  {\ul 0.1415} &
  0.1605 \\
NGCF        & 0.2012  & 0.2723  & 0.1102 & 0.1288  & 0.1742       & 0.2332       & 0.1068       & 0.1249       & 0.2356  & 0.3448       & 0.1286 & 0.1573       \\
LightGCN    & 0.2531  & 0.3561  & 0.1385 & 0.1839  & 0.2585       & {\ul 0.3293} & 0.1479       & 0.1636       & 0.2388  & {\ul 0.3509} & 0.1294 & {\ul 0.1611} \\
SGL         & 0.2586  & 0.3075  & 0.1454 & 0.1576  & {\ul 0.2650} & 0.3210       & {\ul 0.1563} & {\ul 0.1704} & 0.2221  & 0.3030       & 0.1230 & 0.1437       \\ 
CFAG &
  \textbf{0.3540} &
  \textbf{0.4484} &
  \textbf{0.1959} &
  \textbf{0.2184} &
  \textbf{0.4264} &
  \textbf{0.5323} &
  \textbf{0.2590} &
  \textbf{0.2790} &
  \textbf{0.2851} &
  \textbf{0.3856} &
  \textbf{0.1599} &
  \textbf{0.1824} \\ \hline
Improvement & 16.52\% & 25.67\% & 8.53\% & 11.49\% & 60.87\%      & 61.65\%      & 65.71\%      & 63.73\%      & 17.62\% & 9.88\%       & 13.00\% & 13.22\%      \\ \hline
\end{tabular}
\end{table*}

\subsubsection{Datasets} 
\begin{table}\caption{The statistics of datasets.}\label{tab:dataset}
\begin{tabular}{l|l|l|l}
\hline
\hline
\textbf{Dataset}             & \textbf{Mafengwo} & \textbf{Weeplaces} & \textbf{Steam} \\ \hline
\text{\# users}            & 1,269             & 1,501              & 11,099         \\ 
\text{\# groups}           & 972               & 4,651              & 1,085          \\ 
\text{\# items}            & 999               & 6,406              & 2,351          \\ 
\text{\# user-group edges} & 5,574             & 12,258             & 57,654         \\ 
\text{\# user-item edges} & 8,676             & 43,942             & 444,776        \\ 
\text{\# group-item edges} & 2,540             & 6,033              & 22,318         \\ 
\text{Avg. \# groups/user} & 4.39              & 8.17               & 5.19           \\ 
\text{Avg. \# items/user}  & 6.84              & 29.28              & 40.07          \\ 
\text{Avg. \# items/group} & 2.61              & 1.29               & 20.57          \\ \hline
\end{tabular}
\end{table}
We conduct on three real-world datasets: Mafengwo, Weeplaces and Steam. Both Mafengwo and Weeplaces
datasets contain the user’s travel history with a location-based
social network. The history of creating or joining group travel for a user is recorded in Mafengwo~\cite{agree18}. For Weeplaces, we construct group interactions with venues by check-in time and users' social networks in the same way as GroupIM~\cite{groupim@20}.
Both Mafengwo and Weeplaces have limited users as shown in Table~\ref{tab:dataset}. Therefore, we release a new dataset Steam which includes 11,099 users and $57,654$ group participation records on Steam online game platform. 
The statistics of
three datasets is shown in Table 
\ref{tab:dataset}. 
For Mafengwo, we randomly select $70\%$ of all groups joined by each user for training and validation  and the remaining $30\%$ for testing. 
For Weeplaces and Steam, the split ratio is $80\%$ for training and validation and $20\%$ for the testing set. Our implementation is available online\footnote{https://github.com/mdyfrank/CFAG}.

\subsubsection{Baselines}
To justify the effectiveness of our work, we compare the following baselines:
\begin{itemize}[leftmargin=*]
    \item \textbf{AGREE}~\cite{agree18}. This model is designed to recommend items to groups and users, which integrates the user, item and group information. We adapt it to RGI task by endowing AGREE with user-group pairwise BPR loss instead of the original item prediction loss.
    \item \textbf{MF-BPR}~\cite{MF09}. This is the classical pair-wise matrix factorization based recommendation model optimized by the BPR loss.
    \item \textbf{ENMF}~\cite{enmf20}. This model based on a neural matrix factorization architecture leverages mathematical optimization to train the model efficiently without sampling data.
    \item \textbf{NGCF}~\cite{ngcf19}. This method is a variant of standard GCN~\cite{GCN16} leveraging high-order connectivity in a user-item bipartite graph for collaborative filtering.
    \item \textbf{LightGCN}~\cite{lightgcn20}. This is a method based on NGCF with optimization in training efficiency and generation ability by removing feature transformation and nonlinear activation.
    \item \textbf{SGL}~\cite{sgl21}. This work performs contrastive learning on LightGCN to augment node representations for user-item recommendation.
\end{itemize}
Since those user-item recommendation baseline methods are not designed for tripartite graphs, we deploy them with only user-group interactions such that they are adapted to RGI task.

\subsubsection{Parameter Settings}
We apply a grid search for hyperparameters tuning in our model. We searched
embedding size in $\{128$, $256$, $512$, $1024$, $2048\}$, learning rate in $\{0.0001$,
$0.0005$, $0.001$, $0.005$, $0.01\}$, regularization parameter $\lambda_\Theta$ in $\{0.001$, $0.005$, $0.01$, $0.05$, $0.1\}$, and the hyperparameter $\lambda_\alpha$ to control the strength of attention in $\{0.01$, $0.05$, $0.1$, $0.5$, $1\}$. 
We set both personalized embeddings and contextualized embeddings in the same embedding size and leave the exploration of different embedding sizes in future work. A simple division over the dimension is used as partition layer, and direct concatenation is used as merging layer. We use one convolutional layer with batch size $=2048$ for Mafengwo and Weeplaces, and two convolutional layers with batch size $=8196$ for the larger Steam dataset. Early stopping is utilized in all experiments to cope with the over-fitting problem.
\subsubsection{Evaluation Metrics}
We evaluate RGI task by ranking the test groups with all non-interacted groups of users. 
And we adopt Recall$@ \{10,20\}$ and NDCG$@\{10,20\}$ as evaluation metrics.

\subsection{Overall Performance Comparison}
We present the overall comparison results in Table~\ref{tab:performance}. 
The best results among all methods are in boldface, and the second best results are underlined.
We summarize the following key observations:
\begin{itemize}[leftmargin=*]
    \item The proposed CFAG method achieves the best results on all three datasets. Especially, it outperforms all the baseline methods significantly in Weeplaces dataset by more than $60\%$. We hypothesize these large gains result from the abundant user-group interactions as Table 1 shows.
    This demonstrates that CFAG is able to well characterize the user-group interactions.
    The performance gain in the other two datasets is from $8.53\%$ to $25.67\%$, which demonstrates the superiority of the proposed framework.
    \item  The bipartite graph convolutional networks, such as LightGCN and NGCF, are better than a simple matrix factorization method MF-BPR on all the datasets, which indicates the benefits of using graph propagation to learn embeddings.  
    However, they are still unable to incorporate the tripartite graph information, thus being worse than CFAG. 
    This observation justifies the necessity of tripartite graph convolution for RGI task.
    \item Although AGREE integrates both the user-item and user-group interaction information, its poor performance compared to other baseline methods indicates that such methods designed for group-item recommendation tasks cannot be directly applied to RGI task.
    Compared with it, CFAG is specifically designed for RGI task with tripartite graph convolution, and propagation augmentation via factorized attention, which is a better framework.
\end{itemize}

\subsection{Ablation study}\label{sec:ablation}
In this section, we conduct two types of ablation study to justify the effectiveness of those modules in CFAG, which are partition/merging layers in tripartite graph convolution and PA layers.
\subsubsection{Partition and Merging Layer Settings}
\begin{figure}[h]
    \begin{subfigure}{0.155\textwidth}
    \includegraphics[width=\textwidth]{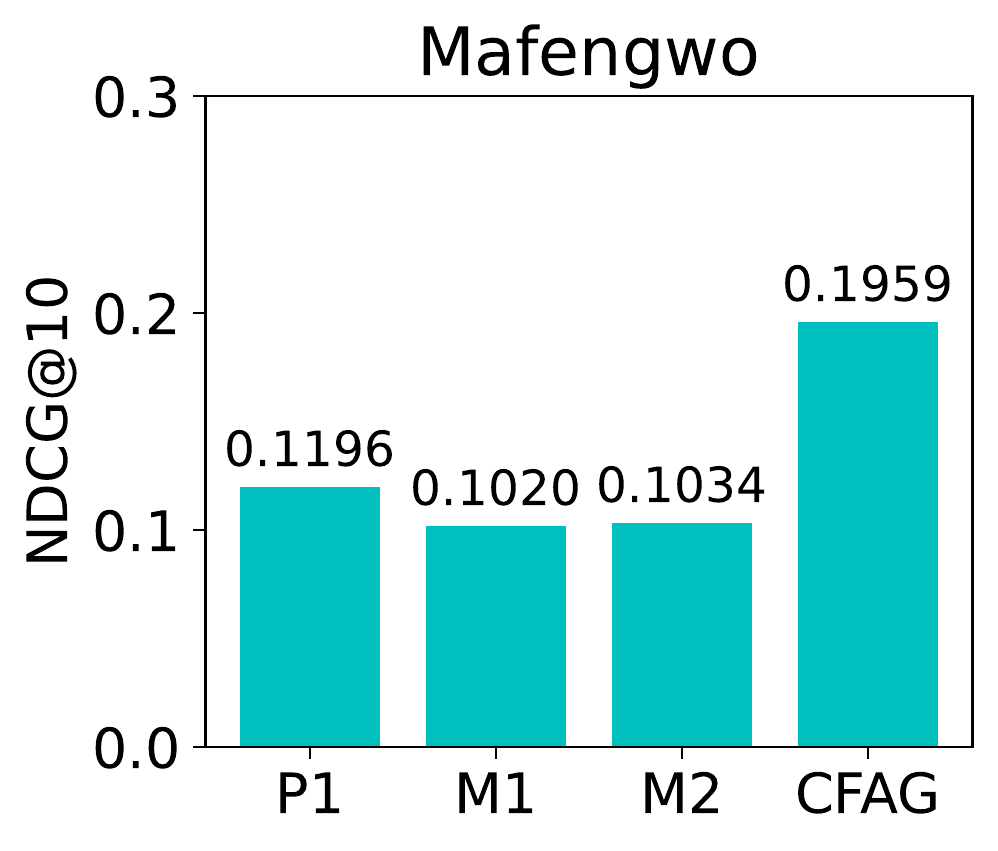}
    \end{subfigure}
    \hfill
    \begin{subfigure}{0.155\textwidth}
    \includegraphics[width=\textwidth]{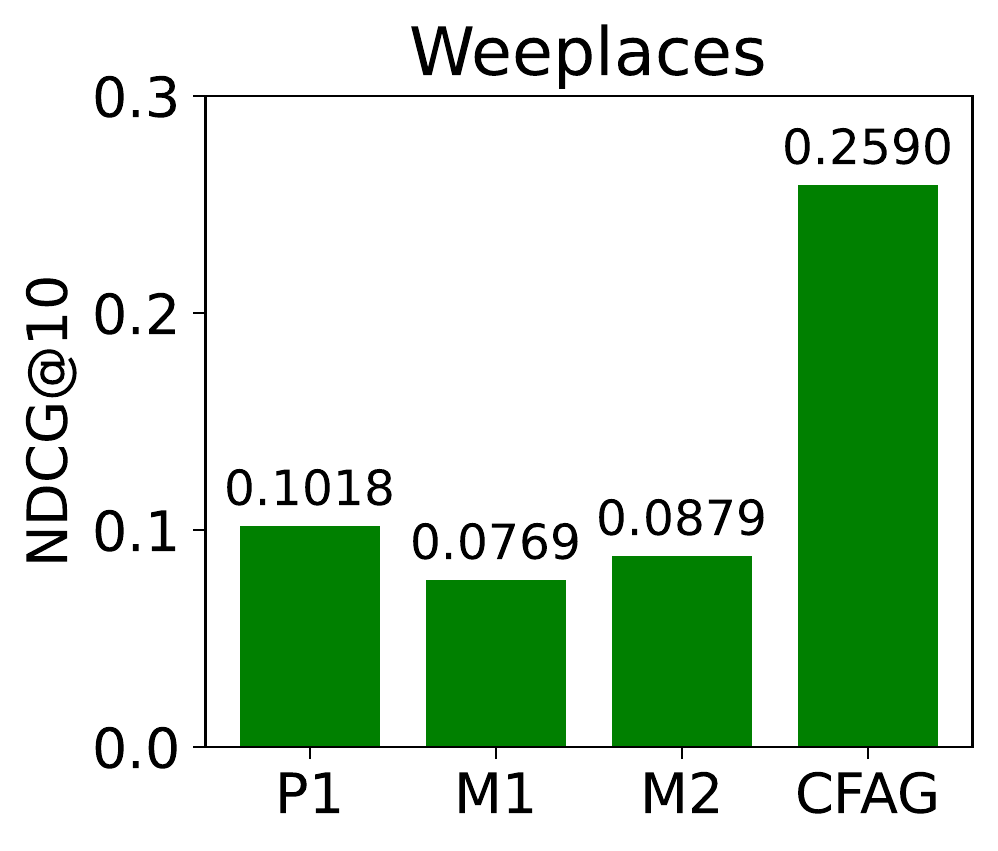}
    \end{subfigure}
    \hfill
    \begin{subfigure}{0.155\textwidth}
    \includegraphics[width=\textwidth]{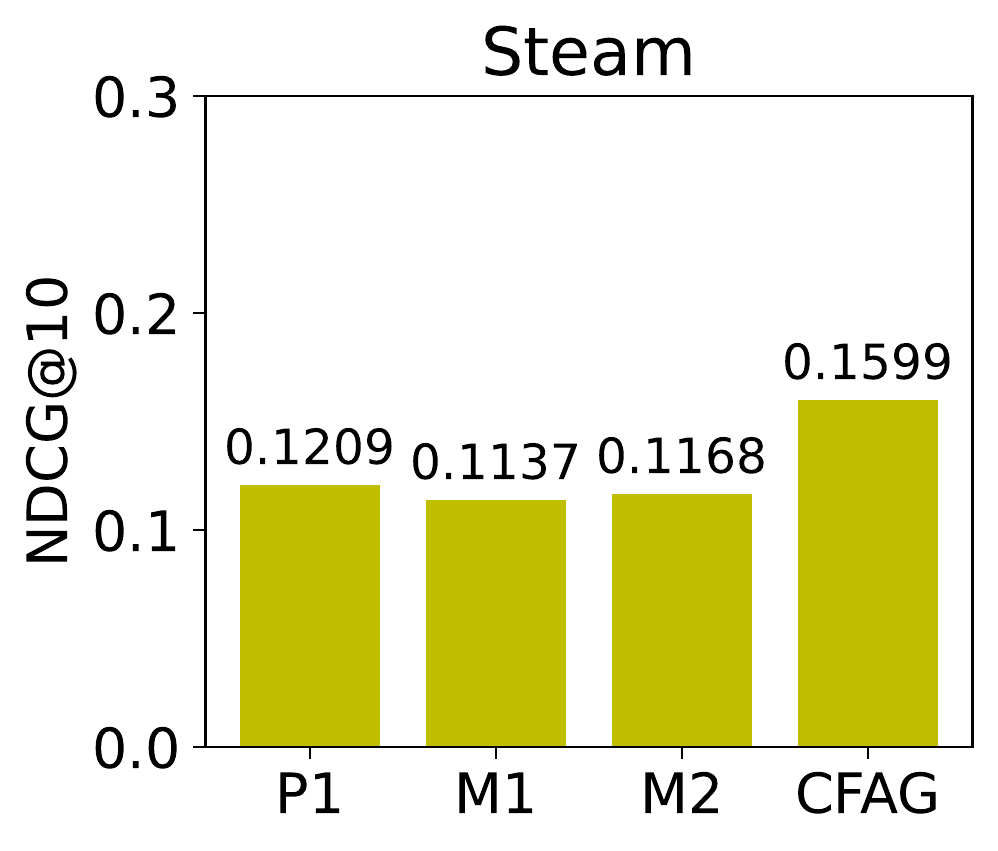}
    \end{subfigure}
        \caption{Performance of CFAG w.r.t. partition/merging layer settings.}
  \label{fig:partition}
\end{figure}

We demonstrate the performance of CFAG with different partition and merging layer settings on the three datasets in Figure~\ref{fig:partition}. 
CFAG employs a simple division over the dimension as partition layer and a direct concatenation as merging layer.
In addition, we investigate three other settings: (P1) Fully-connected layer as partition layer; (M1) concatenation before a fully-connected layer as merging layer; and (M2) concatenation after a fully-connected layer as merging layer. 
We only show the result on NDCG since the pattern on Recall is the same. 
We observe that CFAG consistently performs the best 
on three datasets. 
This justifies the superiority of CFAG with a simple division as partition layer and a direct concatenation as merging layer performs best among all the settings. 
The reason is that those fully connected layers increase the complexity of the model with redundant parameters and also degrade model effectiveness.

\subsubsection{PA layer variants}
Another ablation study is further made to investigate layers.
We design four variants of PA layers. 1) \textit{w/o PA} is without using PA layer in CFAG, which only employs tripartite graph convolution layer. 
2) \textit{w/o group} is the PA layers without the user-to-group propagation augmentation. 
3) \textit{w/o item} is the PA layers without the item-to-user propagation augmentation.
And 4) \textit{GAT att.} is changing the factorized attention mechanism to GAT attention layers in~\cite{gat17}.  We demonstrate the results in Table~\ref{tab:ablation}.

\begin{table}[h]\caption{Ablation study on PA layers.}\label{tab:ablation}
\resizebox{0.47\textwidth}{!}{%
\begin{tabular}{lcccccc}
\hline
Dataset & \multicolumn{2}{c}{Mafengwo}      & \multicolumn{2}{c}{Weeplaces}     & \multicolumn{2}{c}{Steam}         \\ \hline
Metric    & R@10   & N@10   & R@10   & N@10   & R@10   & N@10   \\ \hline
w/o PA  & 0.3338 & 0.1881 & 0.4065 & 0.2419 & 0.2620 & 0.1389 \\
w/o item  & 0.3379 & 0.1913 & 0.4116 & 0.2438 & 0.2838 & 0.1560 \\
w/o group & 0.3381 & 0.1899 & 0.4198 & 0.2504 & 0.2835 & 0.1538 \\
GAT att.  & 0.3190 & 0.1747 & 0.4172 & 0.2484 & 0.2800 & 0.1520 \\ 
\hline
CFAG    & \textbf{0.3540} & \textbf{0.1959} & \textbf{0.4264} & \textbf{0.2590} & \textbf{0.2851} & \textbf{0.1599} \\ \hline
\end{tabular}%
}
\end{table}

From the results, we can observe that without the PA layer, the performance is the worst on both Weeplaces and Steam datasets. 
This verifies the effectiveness of PA layers.
Moreover, if we use only group-to-user or item-to-user PA, the performance is improved compared with w/o PA, which indicates the benefits of PA layers. 
However, w/o item and w/o group PA are both worse than CFAG which uses both PA. This also justifies the necessity of augmenting both user-item interactions and user-group participation.
Additionally, we observe that using the GAT layer to learn attention weights is not able to improve the performance. GAT att. even yields the worst performance on Mafengwo dataset.
We argue that GAT layer introduces redundant parameters to learn the attention weights.
Therefore, we believe that the proposed factorized attention mechanism is a better way to infer the attention weights for PA layers.

\subsection{Contextual Embedding Analysis}
As aforementioned, contextual embeddings characterize the similarity of groups and items, and thus we can construct augmented propagation paths.
To verify this, we conduct analyses of those learned contextual embeddings of CFAG from two perspectives. 

\begin{figure}[h]
  \begin{subfigure}{0.155\textwidth}
    \includegraphics[width=\textwidth]{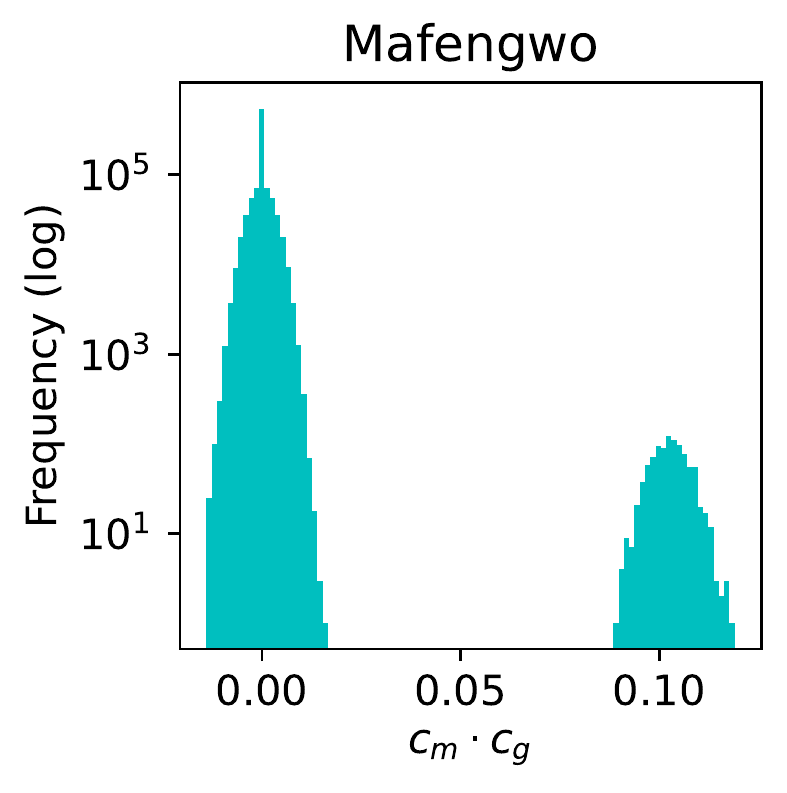}
    \end{subfigure}
    \hfill
    \begin{subfigure}{0.155\textwidth}
    \includegraphics[width=\textwidth]{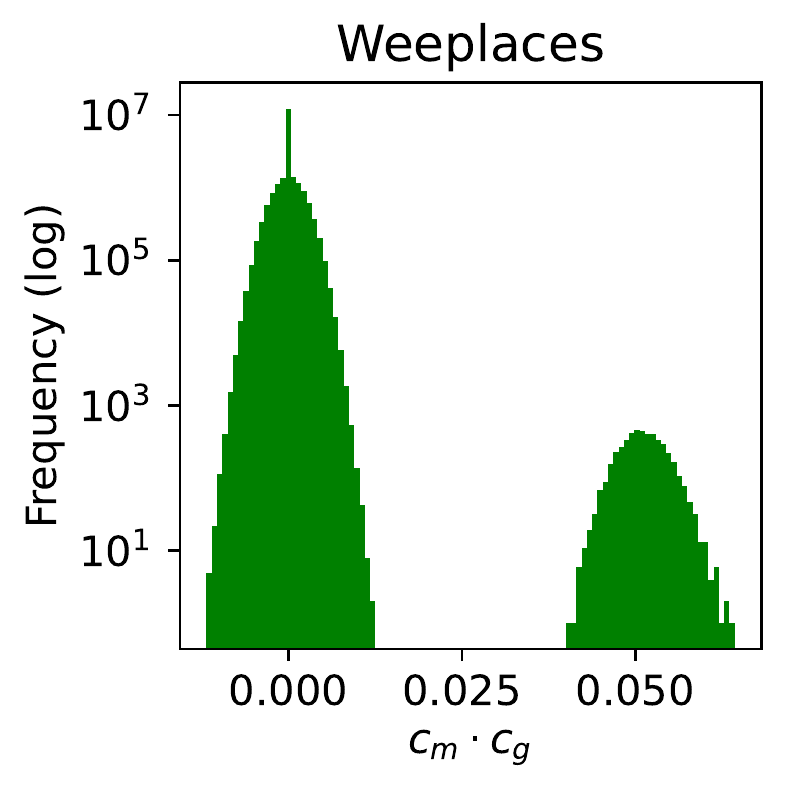}
    \end{subfigure}
    \hfill
    \begin{subfigure}{0.155\textwidth}
    \includegraphics[width=\textwidth]{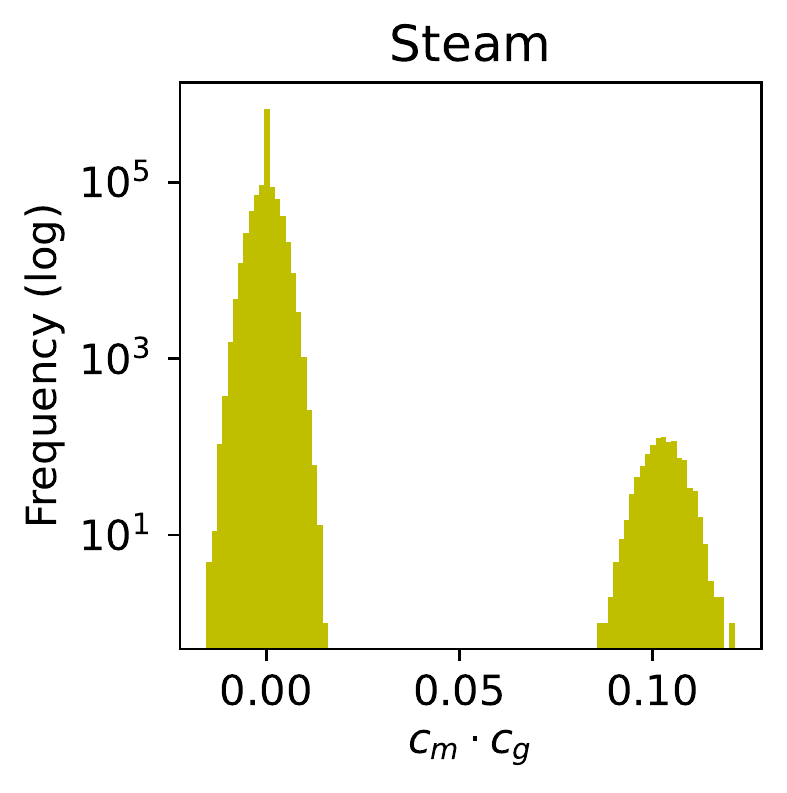}
    \end{subfigure}
    \caption{Distribution of the dot-product values on group contextualized embeddings, \textit{i.e.,} $c_m\cdot c_g$.}
  \label{fig:relatedness of contextual embedding}
\end{figure}

Firstly, we investigate the distribution of all the values in the relatedness matrix $\mathbf{R}$. 
Since values in it are too small, we instead present the pair-wise value before passing to softmax, \textit{i.e.}, the $\mathbf{c}_m \cdot \mathbf{c}_g$. 
Also, due to the space limitation, we only present the results regarding group contextual embeddings. 
Item contextual embeddings have similar patterns.
The distributions on three datasets are present in Figure~\ref{fig:relatedness of contextual embedding}. 
Two obvious peaks appear on all datasets.
The first peak centers at $0$ and is much higher than the second peak, which suggests that most of the groups are not related based on contextual embedding.
The second peak centers at a distinct value.
The second peak is actually those correlated groups and those similar groups constructs augmented propagation paths from groups to users.  
Therefore, we conclude that the learned contextual embedding is able to reveal the similarity of groups and benefits the propagation augmentations. 

\begin{figure}[h]
  \begin{subfigure}{0.16\textwidth}
    \includegraphics[width=\textwidth]{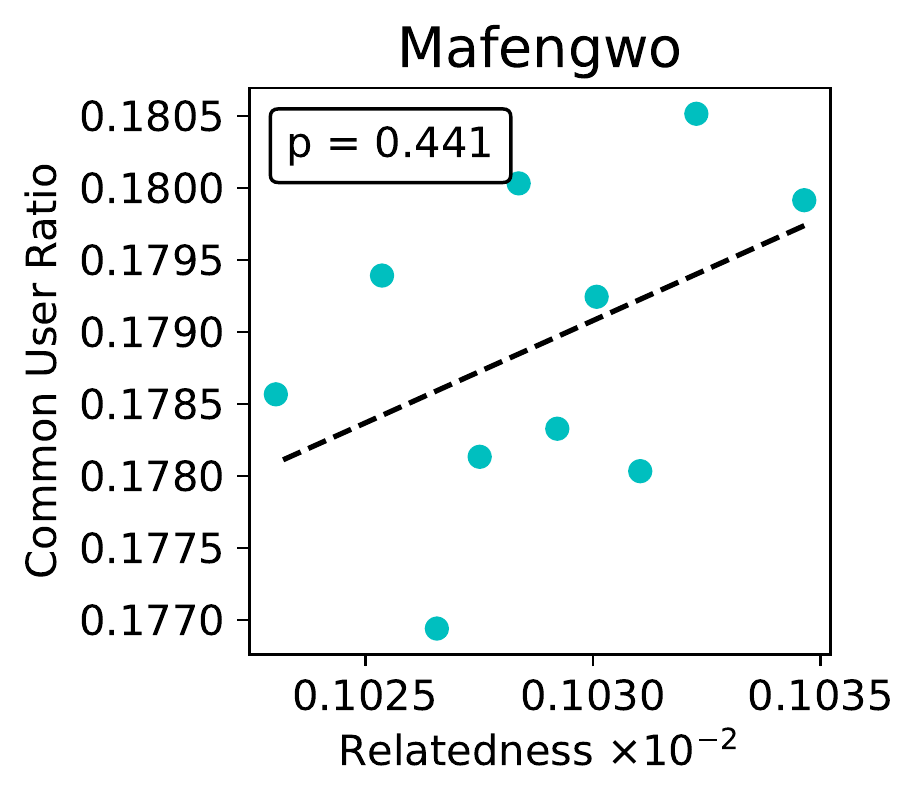}
    \end{subfigure}
    \hfill
    \begin{subfigure}{0.15\textwidth}
    \includegraphics[width=\textwidth]{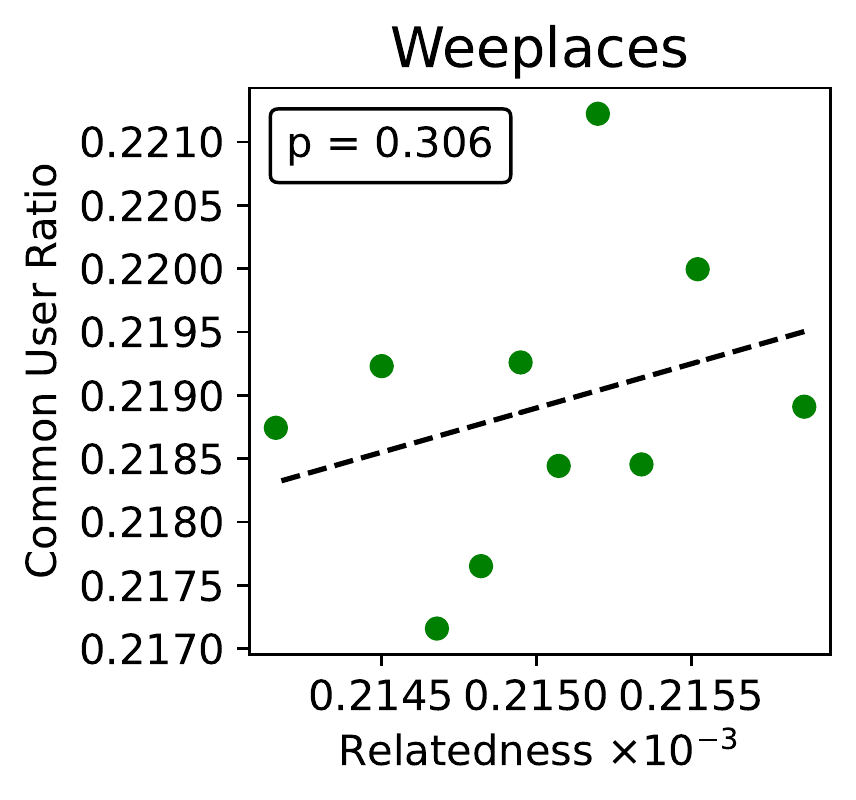}
    \end{subfigure}
    \hfill
    \begin{subfigure}{0.15\textwidth}
    \includegraphics[width=\textwidth]{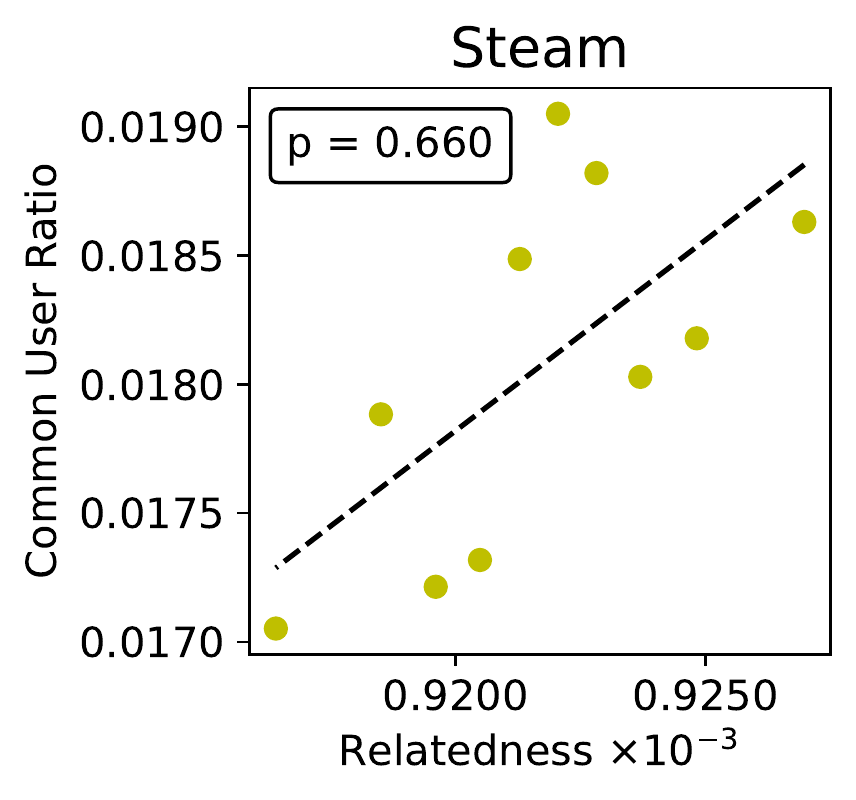}
    \end{subfigure}
    \caption{Common user ratio w.r.t. relatedness of group pairs based on contextual embeddings.}
  \label{fig:common user ration}
\end{figure}

Secondly, 
we retrieve all pairs of groups. Then, we calculate their relatedness and corresponding common user ratio. The ratio for each group pair $(g_a,g_b)$ is computed as follows:
\begin{equation}\label{eq:common user ratio}
r_{ab} = \frac{|\mathcal{N}^{(u)}_{g_a}\cap\mathcal{N}^{(u)}_{g_b}|}{|\mathcal{N}_{g_a}^{(u)}\cup\mathcal{N}^{(u)}_{g_b}|},
\end{equation}
where $\mathcal{N}_{g_a}^{(u)}$ and $\mathcal{N}_{g_b}^{(u)}$ is the set of users in group $g_a$ and $g_b$, respectively. 
Hence, the nominator in Eq.(\ref{eq:common user ratio}) denotes the common users for group pair $(g_a, g_b)$ while the denominator denotes the total number of users in this pair.
For a simple illustration purpose, we sort all group pairs w.r.t. the relatedness scores and split them into $10$ equal size subsets, and represent each subset as the average relatedness scores. Then, we calculate the average common user ratio in each subset.
The scatter plots between relatedness and common user ratio sharing ratio on three datasets  are shown in Figure~\ref{fig:common user ration}. 
We also draw a regression line and compute the Pearson correlation coefficient $p$ for each dataset.  

We have the following observations: Firstly, an overall tendency is that the common user ratio  increases with the growing of relatedness, especially in Mafengwo and Steam datasets.
This tendency indicates that two groups sharing more common members can have higher relatedness based on contextual embeddings, which justifies the efficacy of learned contextual embeddings. 
Secondly, the highest $p$ value in the largest Steam dataset implies that the propagation augmentation ability of contextual embeddings can be more effective on larger datasets. 
\subsection{Cold-start Group Recommendation}
\begin{figure}
    \begin{subfigure}{0.23\textwidth}
    \includegraphics[width=\textwidth]{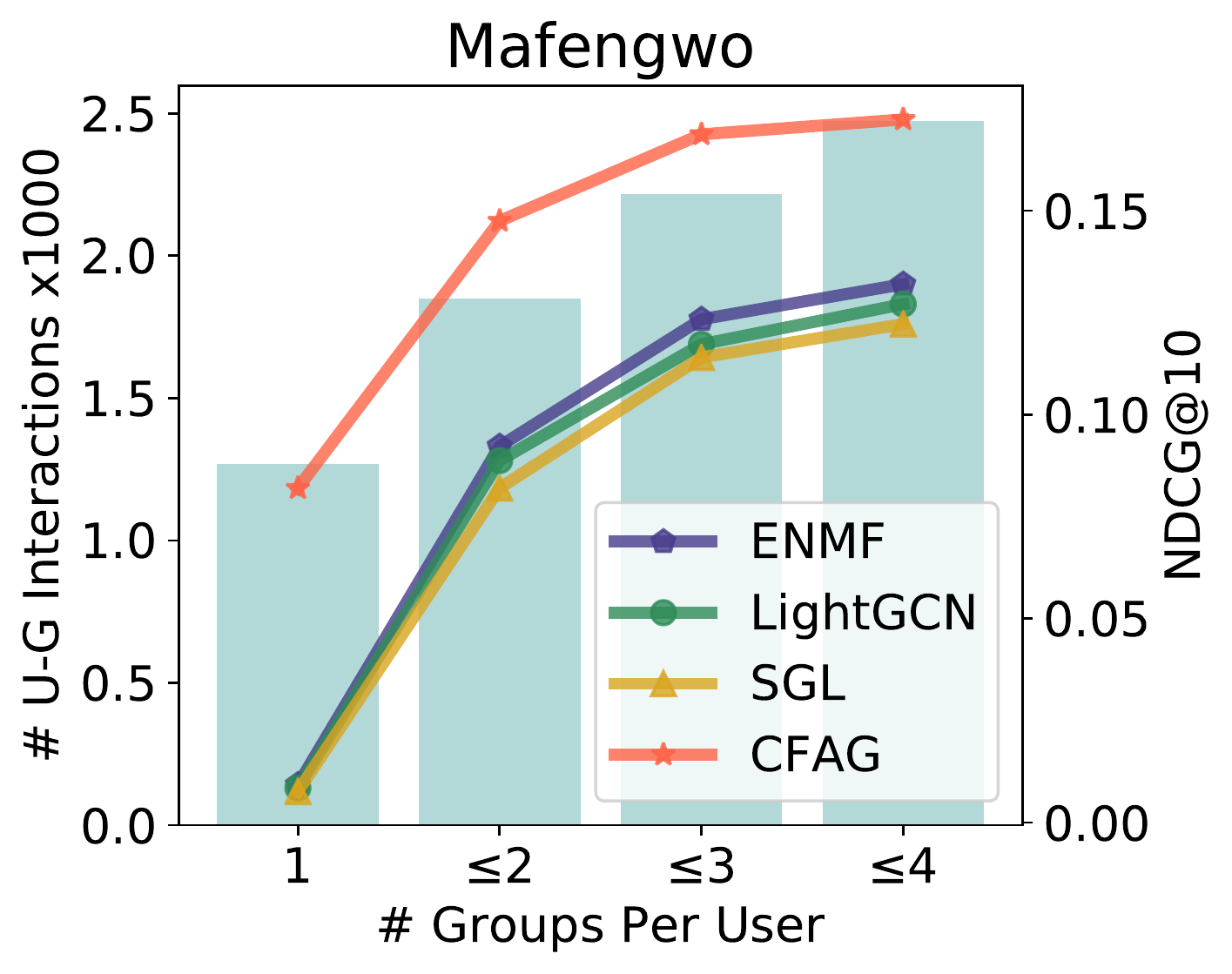}
    \label{fig:mafengwo_cold}
    \end{subfigure}
    \hspace{-2mm}
    \begin{subfigure}{0.23\textwidth}
    \includegraphics[width=\textwidth]{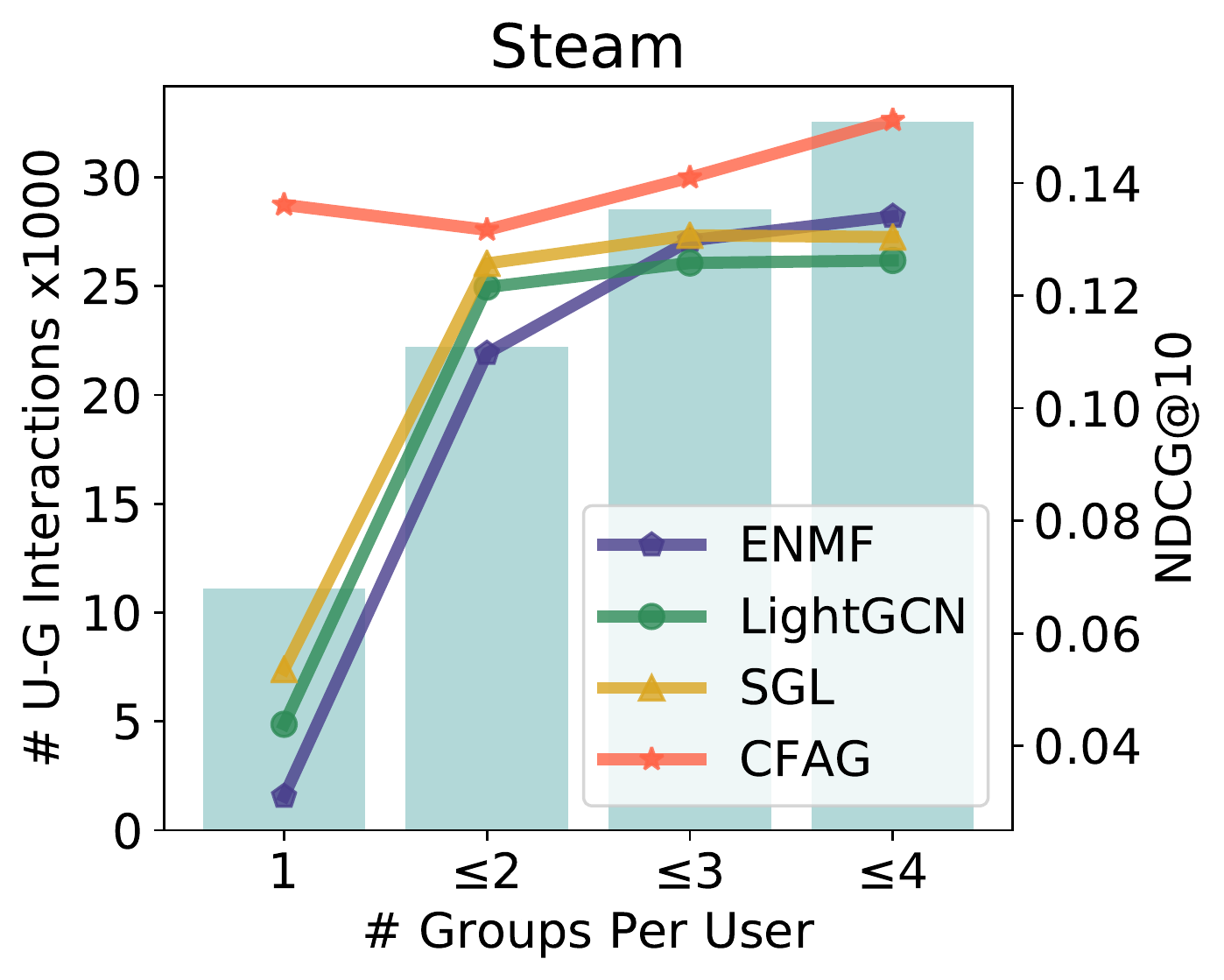}
    \label{fig:Steam_cold}
    \end{subfigure}
    \caption{Cold-start performance of different methods.}
    \label{fig:cold}
\end{figure}

As aforementioned, the cold-start issue in RGI is severe.
Hence, we conduct a detailed analysis regarding the ability of CFAG to tackle cold-start group recommendation. 
We randomly remove some user-group edges for each user in the training set such that the number of neighbor groups of each user is no greater than a threshold $k$. 
For comparison, we choose ENMF, LightGCN, and SGL as the three baseline models, and perform the experiments with threshold $k\in\{1,2,3,4\}$. The threshold indicates the maximum number of groups per user.
The results are in Figure~\ref{fig:cold}.
We report the NDCG performance with respect to different thresholds as the solid line and the number of user-group interactions in the background
histograms.
We observe that CFAG significantly outperforms other baselines.
The reasons are twofold. First, CFAG can leverage both the item and group interactions to learn node embeddings. Though few group participation of users, item interactions complement the cold-start issue. The other reason is that the PA layers in CFAG resolve the cold-start issue by discovering potential unconnected neighbors, thus being able to improve the cold-start performance.

Additionally, we observe that on the large-scale Steam datasets, CFAG even performs better when $k=1$ than on $k=2$. 
We hypothesize that when the number of groups per user is few, CFAG can well characterize the group interests of users from their item interactions. 
Therefore, it verifies the CFAG is a better framework to comprehensively integrate item and group information for users and can successfully complete the RGI task.




\section{Conclusion}
In this paper, we formulate the Ranking-based Group Identification~(RGI) problem, with the goal of recommending new groups to a target user based on the user's group and item interactions. 
In RGI problem, it is challenging to effectively harness both item interactions and group participation on a social tripartite graph.
To address these challenges, we propose a novel framework CFAG, which is able to (i) effectively aggregate information from different types of neighborhoods via the tripartite graph convolution, and (ii) augment the propagation paths to resolve the data sparsity issue via PA layers. 
We conduct extensive experiments and detailed analyses on three datasets to verify the effectiveness of CFAG. 
In the future, we may explore how to select the optimal neighbors for aggregation and different aggregator layers. As such, we can improve the functionality of tripartite graph convolution layers.
\section{Acknowledgements}
This work is supported by National Key R\&D Program of China through grant 2022YFB3104700, NSF under grants III-1763325, III-1909323, III-2106758, SaTC-1930941, S\&T Program of Hebei through grant 21340301D, and Beijing Natural Science Foundation through grant 4222030.

\bibliographystyle{ACM-Reference-Format}
\bibliography{CIKM_Group_Identification}


\begin{thebibliography}{38}


\ifx \showCODEN    \undefined \def \showCODEN     #1{\unskip}     \fi
\ifx \showDOI      \undefined \def \showDOI       #1{#1}\fi
\ifx \showISBNx    \undefined \def \showISBNx     #1{\unskip}     \fi
\ifx \showISBNxiii \undefined \def \showISBNxiii  #1{\unskip}     \fi
\ifx \showISSN     \undefined \def \showISSN      #1{\unskip}     \fi
\ifx \showLCCN     \undefined \def \showLCCN      #1{\unskip}     \fi
\ifx \shownote     \undefined \def \shownote      #1{#1}          \fi
\ifx \showarticletitle \undefined \def \showarticletitle #1{#1}   \fi
\ifx \showURL      \undefined \def \showURL       {\relax}        \fi
\providecommand\bibfield[2]{#2}
\providecommand\bibinfo[2]{#2}
\providecommand\natexlab[1]{#1}
\providecommand\showeprint[2][]{arXiv:#2}

\bibitem[Cao et~al\mbox{.}(2018)]%
        {agree18}
\bibfield{author}{\bibinfo{person}{Da Cao}, \bibinfo{person}{Xiangnan He},
  \bibinfo{person}{Lianhai Miao}, \bibinfo{person}{Yahui An},
  \bibinfo{person}{Chao Yang}, {and} \bibinfo{person}{Richang Hong}.}
  \bibinfo{year}{2018}\natexlab{}.
\newblock \showarticletitle{Attentive Group Recommendation}. In
  \bibinfo{booktitle}{\emph{The 41st International ACM SIGIR Conference on
  Research and Development in Information Retrieval}} (Ann Arbor, MI, USA)
  \emph{(\bibinfo{series}{SIGIR '18})}. \bibinfo{publisher}{Association for
  Computing Machinery}, \bibinfo{address}{New York, NY, USA},
  \bibinfo{pages}{645–654}.
\newblock
\showISBNx{9781450356572}


\bibitem[Chen et~al\mbox{.}(2020)]%
        {enmf20}
\bibfield{author}{\bibinfo{person}{Chong Chen}, \bibinfo{person}{Min Zhang},
  \bibinfo{person}{Yongfeng Zhang}, \bibinfo{person}{Yiqun Liu}, {and}
  \bibinfo{person}{Shaoping Ma}.} \bibinfo{year}{2020}\natexlab{}.
\newblock \showarticletitle{Efficient Neural Matrix Factorization without
  Sampling for Recommendation}.
\newblock \bibinfo{journal}{\emph{ACM Trans. Inf. Syst.}} \bibinfo{volume}{38},
  \bibinfo{number}{2}, Article \bibinfo{articleno}{14} (\bibinfo{date}{Jan}
  \bibinfo{year}{2020}), \bibinfo{numpages}{28}~pages.
\newblock
\showISSN{1046-8188}


\bibitem[Chen et~al\mbox{.}(2008)]%
        {chen08}
\bibfield{author}{\bibinfo{person}{Wen-Yen Chen}, \bibinfo{person}{Dong Zhang},
  {and} \bibinfo{person}{Edward~Y. Chang}.} \bibinfo{year}{2008}\natexlab{}.
\newblock \showarticletitle{Combinational Collaborative Filtering for
  Personalized Community Recommendation}. In
  \bibinfo{booktitle}{\emph{Proceedings of the 14th ACM SIGKDD International
  Conference on Knowledge Discovery and Data Mining}} (Las Vegas, Nevada, USA)
  \emph{(\bibinfo{series}{KDD '08})}. \bibinfo{publisher}{Association for
  Computing Machinery}, \bibinfo{address}{New York, NY, USA},
  \bibinfo{pages}{115–123}.
\newblock
\showISBNx{9781605581934}
\urldef\tempurl%
\url{https://doi.org/10.1145/1401890.1401909}
\showDOI{\tempurl}


\bibitem[Cui et~al\mbox{.}(2019)]%
        {grurs_3}
\bibfield{author}{\bibinfo{person}{Zeyu Cui}, \bibinfo{person}{Zekun Li},
  \bibinfo{person}{Shu Wu}, \bibinfo{person}{Xiao-Yu Zhang}, {and}
  \bibinfo{person}{Liang Wang}.} \bibinfo{year}{2019}\natexlab{}.
\newblock \showarticletitle{Dressing as a Whole: Outfit Compatibility Learning
  Based on Node-Wise Graph Neural Networks}. In \bibinfo{booktitle}{\emph{The
  World Wide Web Conference}} (San Francisco, CA, USA)
  \emph{(\bibinfo{series}{WWW '19})}. \bibinfo{publisher}{Association for
  Computing Machinery}, \bibinfo{address}{New York, NY, USA},
  \bibinfo{pages}{307–317}.
\newblock
\showISBNx{9781450366748}


\bibitem[Ding et~al\mbox{.}(2017)]%
        {fr_2}
\bibfield{author}{\bibinfo{person}{Daizong Ding}, \bibinfo{person}{Mi Zhang},
  \bibinfo{person}{Shao-Yuan Li}, \bibinfo{person}{Jie Tang},
  \bibinfo{person}{Xiaotie Chen}, {and} \bibinfo{person}{Zhi-Hua Zhou}.}
  \bibinfo{year}{2017}\natexlab{}.
\newblock \showarticletitle{BayDNN: Friend Recommendation with Bayesian
  Personalized Ranking Deep Neural Network}. In
  \bibinfo{booktitle}{\emph{Proceedings of the 2017 ACM on Conference on
  Information and Knowledge Management}} (Singapore, Singapore)
  \emph{(\bibinfo{series}{CIKM '17})}. \bibinfo{publisher}{Association for
  Computing Machinery}, \bibinfo{address}{New York, NY, USA},
  \bibinfo{pages}{1479–1488}.
\newblock
\showISBNx{9781450349185}


\bibitem[Fan et~al\mbox{.}(2019a)]%
        {graphRec}
\bibfield{author}{\bibinfo{person}{Wenqi Fan}, \bibinfo{person}{Yao Ma},
  \bibinfo{person}{Qing Li}, \bibinfo{person}{Yuan He}, \bibinfo{person}{Eric
  Zhao}, \bibinfo{person}{Jiliang Tang}, {and} \bibinfo{person}{Dawei Yin}.}
  \bibinfo{year}{2019}\natexlab{a}.
\newblock \showarticletitle{Graph Neural Networks for Social Recommendation}.
  In \bibinfo{booktitle}{\emph{The World Wide Web Conference}} (San Francisco,
  CA, USA) \emph{(\bibinfo{series}{WWW '19})}. \bibinfo{publisher}{Association
  for Computing Machinery}, \bibinfo{address}{New York, NY, USA},
  \bibinfo{pages}{417–426}.
\newblock
\showISBNx{9781450366748}


\bibitem[Fan et~al\mbox{.}(2019b)]%
        {sr_rw1}
\bibfield{author}{\bibinfo{person}{Wenqi Fan}, \bibinfo{person}{Yao Ma},
  \bibinfo{person}{Dawei Yin}, \bibinfo{person}{Jianping Wang},
  \bibinfo{person}{Jiliang Tang}, {and} \bibinfo{person}{Qing Li}.}
  \bibinfo{year}{2019}\natexlab{b}.
\newblock \showarticletitle{Deep Social Collaborative Filtering}. In
  \bibinfo{booktitle}{\emph{Proceedings of the 13th ACM Conference on
  Recommender Systems}} (Copenhagen, Denmark) \emph{(\bibinfo{series}{RecSys
  '19})}. \bibinfo{publisher}{Association for Computing Machinery},
  \bibinfo{address}{New York, NY, USA}, \bibinfo{pages}{305–313}.
\newblock
\showISBNx{9781450362436}


\bibitem[Fortunato(2010)]%
        {fortunato2010community}
\bibfield{author}{\bibinfo{person}{Santo Fortunato}.}
  \bibinfo{year}{2010}\natexlab{}.
\newblock \showarticletitle{Community detection in graphs}.
\newblock \bibinfo{journal}{\emph{Physics reports}} \bibinfo{volume}{486},
  \bibinfo{number}{3-5} (\bibinfo{year}{2010}), \bibinfo{pages}{75--174}.
\newblock


\bibitem[Guo et~al\mbox{.}(2015)]%
        {fr_1}
\bibfield{author}{\bibinfo{person}{Linke Guo}, \bibinfo{person}{Chi Zhang},
  {and} \bibinfo{person}{Yuguang Fang}.} \bibinfo{year}{2015}\natexlab{}.
\newblock \showarticletitle{A Trust-Based Privacy-Preserving Friend
  Recommendation Scheme for Online Social Networks}.
\newblock \bibinfo{journal}{\emph{IEEE Transactions on Dependable and Secure
  Computing}} \bibinfo{volume}{12}, \bibinfo{number}{4} (\bibinfo{year}{2015}),
  \bibinfo{pages}{413--427}.
\newblock


\bibitem[Guo and Wang(2021)]%
        {sr_1}
\bibfield{author}{\bibinfo{person}{Zhiwei Guo} {and} \bibinfo{person}{Heng
  Wang}.} \bibinfo{year}{2021}\natexlab{}.
\newblock \showarticletitle{A Deep Graph Neural Network-Based Mechanism for
  Social Recommendations}.
\newblock \bibinfo{journal}{\emph{IEEE Transactions on Industrial Informatics}}
  \bibinfo{volume}{17}, \bibinfo{number}{4} (\bibinfo{year}{2021}),
  \bibinfo{pages}{2776--2783}.
\newblock


\bibitem[Hamilton et~al\mbox{.}(2017)]%
        {hamilton2017inductive}
\bibfield{author}{\bibinfo{person}{Will Hamilton}, \bibinfo{person}{Zhitao
  Ying}, {and} \bibinfo{person}{Jure Leskovec}.}
  \bibinfo{year}{2017}\natexlab{}.
\newblock \showarticletitle{Inductive representation learning on large graphs}.
\newblock \bibinfo{journal}{\emph{Advances in neural information processing
  systems}}  \bibinfo{volume}{30} (\bibinfo{year}{2017}).
\newblock


\bibitem[Han et~al\mbox{.}(2018)]%
        {han@18}
\bibfield{author}{\bibinfo{person}{Xiaotian Han}, \bibinfo{person}{Chuan Shi},
  \bibinfo{person}{Senzhang Wang}, \bibinfo{person}{Philip~S. Yu}, {and}
  \bibinfo{person}{Li Song}.} \bibinfo{year}{2018}\natexlab{}.
\newblock \showarticletitle{Aspect-Level Deep Collaborative Filtering via
  Heterogeneous Information Networks}. In \bibinfo{booktitle}{\emph{Proceedings
  of the Twenty-Seventh International Joint Conference on Artificial
  Intelligence, {IJCAI-18}}}. \bibinfo{publisher}{International Joint
  Conferences on Artificial Intelligence Organization},
  \bibinfo{pages}{3393--3399}.
\newblock


\bibitem[Hao et~al\mbox{.}(2020)]%
        {hao2020p}
\bibfield{author}{\bibinfo{person}{Junheng Hao}, \bibinfo{person}{Tong Zhao},
  \bibinfo{person}{Jin Li}, \bibinfo{person}{Xin~Luna Dong},
  \bibinfo{person}{Christos Faloutsos}, \bibinfo{person}{Yizhou Sun}, {and}
  \bibinfo{person}{Wei Wang}.} \bibinfo{year}{2020}\natexlab{}.
\newblock \showarticletitle{P-Companion: {A} Principled Framework for
  Diversified Complementary Product Recommendation}. In
  \bibinfo{booktitle}{\emph{{CIKM} '20: The 29th {ACM} International Conference
  on Information and Knowledge Management, Virtual Event, Ireland, October
  19-23, 2020}}. \bibinfo{publisher}{{ACM}}, \bibinfo{pages}{2517--2524}.
\newblock


\bibitem[He et~al\mbox{.}(2020)]%
        {lightgcn20}
\bibfield{author}{\bibinfo{person}{Xiangnan He}, \bibinfo{person}{Kuan Deng},
  \bibinfo{person}{Xiang Wang}, \bibinfo{person}{Yan Li},
  \bibinfo{person}{YongDong Zhang}, {and} \bibinfo{person}{Meng Wang}.}
  \bibinfo{year}{2020}\natexlab{}.
\newblock \bibinfo{booktitle}{\emph{LightGCN: Simplifying and Powering Graph
  Convolution Network for Recommendation}}.
\newblock \bibinfo{publisher}{Association for Computing Machinery},
  \bibinfo{address}{New York, NY, USA}, \bibinfo{pages}{639–648}.
\newblock
\showISBNx{9781450380164}


\bibitem[Hu et~al\mbox{.}(2019)]%
        {Hu19}
\bibfield{author}{\bibinfo{person}{Liang Hu}, \bibinfo{person}{Songlei Jian},
  \bibinfo{person}{Longbing Cao}, \bibinfo{person}{Zhiping Gu},
  \bibinfo{person}{Qingkui Chen}, {and} \bibinfo{person}{Artak Amirbekyan}.}
  \bibinfo{year}{2019}\natexlab{}.
\newblock \showarticletitle{HERS: Modeling Influential Contexts with
  Heterogeneous Relations for Sparse and Cold-Start Recommendation}.
\newblock \bibinfo{journal}{\emph{Proceedings of the AAAI Conference on
  Artificial Intelligence}}  \bibinfo{volume}{33} (\bibinfo{date}{Jul.}
  \bibinfo{year}{2019}), \bibinfo{pages}{3830--3837}.
\newblock


\bibitem[Isinkaye et~al\mbox{.}(2015)]%
        {ISINKAYE@15}
\bibfield{author}{\bibinfo{person}{F.O. Isinkaye}, \bibinfo{person}{Y.O.
  Folajimi}, {and} \bibinfo{person}{B.A. Ojokoh}.}
  \bibinfo{year}{2015}\natexlab{}.
\newblock \showarticletitle{Recommendation systems: Principles, methods and
  evaluation}.
\newblock \bibinfo{journal}{\emph{Egyptian Informatics Journal}}
  \bibinfo{volume}{16}, \bibinfo{number}{3} (\bibinfo{year}{2015}),
  \bibinfo{pages}{261--273}.
\newblock
\showISSN{1110-8665}


\bibitem[Jin et~al\mbox{.}(2019)]%
        {Jin19}
\bibfield{author}{\bibinfo{person}{Di Jin}, \bibinfo{person}{Ziyang Liu},
  \bibinfo{person}{Weihao Li}, \bibinfo{person}{Dongxiao He}, {and}
  \bibinfo{person}{Weixiong Zhang}.} \bibinfo{year}{2019}\natexlab{}.
\newblock \showarticletitle{Graph Convolutional Networks Meet Markov Random
  Fields: Semi-Supervised Community Detection in Attribute Networks}.
\newblock \bibinfo{journal}{\emph{Proceedings of the AAAI Conference on
  Artificial Intelligence}}  \bibinfo{volume}{33} (\bibinfo{date}{Jul.}
  \bibinfo{year}{2019}), \bibinfo{pages}{152--159}.
\newblock


\bibitem[Karimpour et~al\mbox{.}(2021a)]%
        {grouprec21}
\bibfield{author}{\bibinfo{person}{Davod Karimpour}, \bibinfo{person}{Mohammad
  Ali~Zare Chahooki}, {and} \bibinfo{person}{Ali Hashemi}.}
  \bibinfo{year}{2021}\natexlab{a}.
\newblock \showarticletitle{GroupRec: Group Recommendation by Numerical
  Characteristics of Groups in Telegram}. In \bibinfo{booktitle}{\emph{2021
  11th International Conference on Computer Engineering and Knowledge
  (ICCKE)}}. \bibinfo{pages}{115--120}.
\newblock
\urldef\tempurl%
\url{https://doi.org/10.1109/ICCKE54056.2021.9721494}
\showDOI{\tempurl}


\bibitem[Karimpour et~al\mbox{.}(2021b)]%
        {Telegram21}
\bibfield{author}{\bibinfo{person}{Davod Karimpour},
  \bibinfo{person}{Mohammad~Ali Zare~Chahooki}, {and} \bibinfo{person}{Ali
  Hashemi}.} \bibinfo{year}{2021}\natexlab{b}.
\newblock \showarticletitle{Telegram group recommendation based on users'
  migration}. In \bibinfo{booktitle}{\emph{2021 26th International Computer
  Conference, Computer Society of Iran (CSICC)}}. \bibinfo{pages}{1--6}.
\newblock
\urldef\tempurl%
\url{https://doi.org/10.1109/CSICC52343.2021.9420581}
\showDOI{\tempurl}


\bibitem[Kingma and Ba(2015)]%
        {adam14}
\bibfield{author}{\bibinfo{person}{Diederik~P. Kingma} {and}
  \bibinfo{person}{Jimmy Ba}.} \bibinfo{year}{2015}\natexlab{}.
\newblock \showarticletitle{Adam: {A} Method for Stochastic Optimization}. In
  \bibinfo{booktitle}{\emph{3rd International Conference on Learning
  Representations, {ICLR} 2015, San Diego, CA, USA, May 7-9, 2015, Conference
  Track Proceedings}}.
\newblock


\bibitem[Kipf and Welling(2017)]%
        {GCN16}
\bibfield{author}{\bibinfo{person}{Thomas~N. Kipf} {and} \bibinfo{person}{Max
  Welling}.} \bibinfo{year}{2017}\natexlab{}.
\newblock \showarticletitle{Semi-Supervised Classification with Graph
  Convolutional Networks}. In \bibinfo{booktitle}{\emph{5th International
  Conference on Learning Representations, {ICLR} 2017, Toulon, France, April
  24-26, 2017, Conference Track Proceedings}}.
\newblock


\bibitem[Park et~al\mbox{.}(2017)]%
        {park2017deep}
\bibfield{author}{\bibinfo{person}{Keunchan Park}, \bibinfo{person}{Jisoo Lee},
  {and} \bibinfo{person}{Jaeho Choi}.} \bibinfo{year}{2017}\natexlab{}.
\newblock \showarticletitle{Deep Neural Networks for News Recommendations}. In
  \bibinfo{booktitle}{\emph{Proceedings of the 2017 {ACM} on Conference on
  Information and Knowledge Management, {CIKM} 2017, Singapore, November 06 -
  10, 2017}}. \bibinfo{publisher}{{ACM}}, \bibinfo{pages}{2255--2258}.
\newblock


\bibitem[Rendle et~al\mbox{.}(2009)]%
        {MF09}
\bibfield{author}{\bibinfo{person}{Steffen Rendle}, \bibinfo{person}{Christoph
  Freudenthaler}, \bibinfo{person}{Zeno Gantner}, {and} \bibinfo{person}{Lars
  Schmidt-Thieme}.} \bibinfo{year}{2009}\natexlab{}.
\newblock \showarticletitle{BPR: Bayesian Personalized Ranking from Implicit
  Feedback}. In \bibinfo{booktitle}{\emph{Proceedings of the Twenty-Fifth
  Conference on Uncertainty in Artificial Intelligence}} (Montreal, Quebec,
  Canada) \emph{(\bibinfo{series}{UAI '09})}. \bibinfo{publisher}{AUAI Press},
  \bibinfo{address}{Arlington, Virginia, USA}, \bibinfo{pages}{452–461}.
\newblock
\showISBNx{9780974903958}


\bibitem[Sankar et~al\mbox{.}(2020)]%
        {groupim@20}
\bibfield{author}{\bibinfo{person}{Aravind Sankar}, \bibinfo{person}{Yanhong
  Wu}, \bibinfo{person}{Yuhang Wu}, \bibinfo{person}{Wei Zhang},
  \bibinfo{person}{Hao Yang}, {and} \bibinfo{person}{Hari Sundaram}.}
  \bibinfo{year}{2020}\natexlab{}.
\newblock \bibinfo{booktitle}{\emph{GroupIM: A Mutual Information Maximization
  Framework for Neural Group Recommendation}}.
\newblock \bibinfo{publisher}{Association for Computing Machinery},
  \bibinfo{address}{New York, NY, USA}, \bibinfo{pages}{1279–1288}.
\newblock
\showISBNx{9781450380164}


\bibitem[Shimp and Andrews(2012)]%
        {promotion@12}
\bibfield{author}{\bibinfo{person}{Terence~A Shimp} {and}
  \bibinfo{person}{J~Craig Andrews}.} \bibinfo{year}{2012}\natexlab{}.
\newblock \bibinfo{booktitle}{\emph{Advertising promotion and other aspects of
  integrated marketing communications}}.
\newblock \bibinfo{publisher}{Cengage Learning}.
\newblock


\bibitem[Tao et~al\mbox{.}(2020)]%
        {mgat@20}
\bibfield{author}{\bibinfo{person}{Zhulin Tao}, \bibinfo{person}{Yinwei Wei},
  \bibinfo{person}{Xiang Wang}, \bibinfo{person}{Xiangnan He},
  \bibinfo{person}{Xianglin Huang}, {and} \bibinfo{person}{Tat-Seng Chua}.}
  \bibinfo{year}{2020}\natexlab{}.
\newblock \showarticletitle{MGAT: Multimodal Graph Attention Network for
  Recommendation}.
\newblock \bibinfo{journal}{\emph{Information Processing and Management}}
  \bibinfo{volume}{57}, \bibinfo{number}{5} (\bibinfo{year}{2020}),
  \bibinfo{pages}{102277}.
\newblock
\showISSN{0306-4573}


\bibitem[Velickovic et~al\mbox{.}(2018)]%
        {gat17}
\bibfield{author}{\bibinfo{person}{Petar Velickovic}, \bibinfo{person}{Guillem
  Cucurull}, \bibinfo{person}{Arantxa Casanova}, \bibinfo{person}{Adriana
  Romero}, \bibinfo{person}{Pietro Li{\`{o}}}, {and} \bibinfo{person}{Yoshua
  Bengio}.} \bibinfo{year}{2018}\natexlab{}.
\newblock \showarticletitle{Graph Attention Networks}. In
  \bibinfo{booktitle}{\emph{6th International Conference on Learning
  Representations, {ICLR} 2018, Vancouver, BC, Canada, April 30 - May 3, 2018,
  Conference Track Proceedings}}. \bibinfo{publisher}{OpenReview.net}.
\newblock


\bibitem[Wang et~al\mbox{.}(2012)]%
        {wang12}
\bibfield{author}{\bibinfo{person}{Jingdong Wang}, \bibinfo{person}{Zhe Zhao},
  \bibinfo{person}{Jiazhen Zhou}, \bibinfo{person}{Hao Wang},
  \bibinfo{person}{Bin Cui}, {and} \bibinfo{person}{Guojun Qi}.}
  \bibinfo{year}{2012}\natexlab{}.
\newblock \showarticletitle{Recommending Flickr Groups with Social Topic
  Model}.
\newblock \bibinfo{journal}{\emph{Inf. Retr.}} \bibinfo{volume}{15},
  \bibinfo{number}{3–4} (\bibinfo{date}{jun} \bibinfo{year}{2012}),
  \bibinfo{pages}{278–295}.
\newblock
\showISSN{1386-4564}
\urldef\tempurl%
\url{https://doi.org/10.1007/s10791-012-9193-0}
\showDOI{\tempurl}


\bibitem[Wang et~al\mbox{.}(2021)]%
        {GLRSreview}
\bibfield{author}{\bibinfo{person}{Shoujin Wang}, \bibinfo{person}{Liang Hu},
  \bibinfo{person}{Yan Wang}, \bibinfo{person}{Xiangnan He},
  \bibinfo{person}{{Quan Z.} Sheng}, \bibinfo{person}{{Mehmet A.} Orgun},
  \bibinfo{person}{Longbing Cao}, \bibinfo{person}{Francesco Ricci}, {and}
  \bibinfo{person}{{Philip S.} Yu}.} \bibinfo{year}{2021}\natexlab{}.
\newblock \showarticletitle{Graph learning based recommender systems: a
  review}. In \bibinfo{booktitle}{\emph{Proceedings of the 30th International
  Joint Conference on Artificial Intelligence}}.
  \bibinfo{publisher}{International Joint Conferences on Artificial
  Intelligence}, \bibinfo{pages}{4644--4652}.
\newblock
\newblock
\shownote{30th International Joint Conference on Artificial Intelligence, IJCAI
  2021 ; Conference date: 19-08-2021 Through 27-08-2021}.


\bibitem[Wang et~al\mbox{.}(2016)]%
        {wang16}
\bibfield{author}{\bibinfo{person}{Xin Wang}, \bibinfo{person}{Roger
  Donaldson}, \bibinfo{person}{Christopher Nell}, \bibinfo{person}{Peter
  Gorniak}, \bibinfo{person}{Martin Ester}, {and} \bibinfo{person}{Jiajun Bu}.}
  \bibinfo{year}{2016}\natexlab{}.
\newblock \showarticletitle{Recommending Groups to Users Using User-Group
  Engagement and Time-Dependent Matrix Factorization}. In
  \bibinfo{booktitle}{\emph{Proceedings of the Thirtieth AAAI Conference on
  Artificial Intelligence}} (Phoenix, Arizona)
  \emph{(\bibinfo{series}{AAAI'16})}. \bibinfo{publisher}{AAAI Press},
  \bibinfo{pages}{1331–1337}.
\newblock


\bibitem[Wang et~al\mbox{.}(2019)]%
        {ngcf19}
\bibfield{author}{\bibinfo{person}{Xiang Wang}, \bibinfo{person}{Xiangnan He},
  \bibinfo{person}{Meng Wang}, \bibinfo{person}{Fuli Feng}, {and}
  \bibinfo{person}{Tat-Seng Chua}.} \bibinfo{year}{2019}\natexlab{}.
\newblock \showarticletitle{Neural Graph Collaborative Filtering}. In
  \bibinfo{booktitle}{\emph{Proceedings of the 42nd International ACM SIGIR
  Conference on Research and Development in Information Retrieval}} (Paris,
  France) \emph{(\bibinfo{series}{SIGIR'19})}. \bibinfo{publisher}{Association
  for Computing Machinery}, \bibinfo{address}{New York, NY, USA},
  \bibinfo{pages}{165–174}.
\newblock
\showISBNx{9781450361729}


\bibitem[Wen et~al\mbox{.}(2018)]%
        {sr_emb}
\bibfield{author}{\bibinfo{person}{Yufei Wen}, \bibinfo{person}{Lei Guo},
  \bibinfo{person}{Zhumin Chen}, {and} \bibinfo{person}{Jun Ma}.}
  \bibinfo{year}{2018}\natexlab{}.
\newblock \showarticletitle{Network Embedding Based Recommendation Method in
  Social Networks}. In \bibinfo{booktitle}{\emph{Companion Proceedings of the
  The Web Conference 2018}} (Lyon, France) \emph{(\bibinfo{series}{WWW '18})}.
  \bibinfo{publisher}{International World Wide Web Conferences Steering
  Committee}, \bibinfo{address}{Republic and Canton of Geneva, CHE},
  \bibinfo{pages}{11–12}.
\newblock
\showISBNx{9781450356404}


\bibitem[Wu et~al\mbox{.}(2021)]%
        {sgl21}
\bibfield{author}{\bibinfo{person}{Jiancan Wu}, \bibinfo{person}{Xiang Wang},
  \bibinfo{person}{Fuli Feng}, \bibinfo{person}{Xiangnan He},
  \bibinfo{person}{Liang Chen}, \bibinfo{person}{Jianxun Lian}, {and}
  \bibinfo{person}{Xing Xie}.} \bibinfo{year}{2021}\natexlab{}.
\newblock \showarticletitle{Self-supervised Graph Learning for Recommendation}.
  In \bibinfo{booktitle}{\emph{{SIGIR} '21: The 44th International {ACM}
  {SIGIR} Conference on Research and Development in Information Retrieval,
  Virtual Event, Canada, July 11-15, 2021}}. \bibinfo{publisher}{{ACM}},
  \bibinfo{pages}{726--735}.
\newblock


\bibitem[Yang et~al\mbox{.}(2021b)]%
        {yang21}
\bibfield{author}{\bibinfo{person}{Lu Yang}, \bibinfo{person}{Yezheng Liu},
  \bibinfo{person}{Yuanchun Jiang}, \bibinfo{person}{Le Wu}, {and}
  \bibinfo{person}{Jianshan Sun}.} \bibinfo{year}{2021}\natexlab{b}.
\newblock \showarticletitle{Predicting personalized grouping and consumption: A
  collaborative evolution model}.
\newblock \bibinfo{journal}{\emph{Knowledge-Based Systems}}
  \bibinfo{volume}{228} (\bibinfo{year}{2021}), \bibinfo{pages}{107248}.
\newblock
\showISSN{0950-7051}
\urldef\tempurl%
\url{https://doi.org/10.1016/j.knosys.2021.107248}
\showDOI{\tempurl}


\bibitem[Yang et~al\mbox{.}(2021a)]%
        {ConsisRec}
\bibfield{author}{\bibinfo{person}{Liangwei Yang}, \bibinfo{person}{Zhiwei
  Liu}, \bibinfo{person}{Yingtong Dou}, \bibinfo{person}{Jing Ma}, {and}
  \bibinfo{person}{Philip~S. Yu}.} \bibinfo{year}{2021}\natexlab{a}.
\newblock \showarticletitle{ConsisRec: Enhancing {GNN} for Social
  Recommendation via Consistent Neighbor Aggregation}. In
  \bibinfo{booktitle}{\emph{{SIGIR} '21: The 44th International {ACM} {SIGIR}
  Conference on Research and Development in Information Retrieval, Virtual
  Event, Canada, July 11-15, 2021}}. \bibinfo{publisher}{{ACM}},
  \bibinfo{pages}{2141--2145}.
\newblock


\bibitem[Yang et~al\mbox{.}(2022)]%
        {social_3}
\bibfield{author}{\bibinfo{person}{Liangwei Yang}, \bibinfo{person}{Zhiwei
  Liu}, \bibinfo{person}{Yu Wang}, \bibinfo{person}{Chen Wang},
  \bibinfo{person}{Ziwei Fan}, {and} \bibinfo{person}{Philip~S. Yu}.}
  \bibinfo{year}{2022}\natexlab{}.
\newblock \showarticletitle{Large-scale Personalized Video Game Recommendation
  via Social-aware Contextualized Graph Neural Network}. In
  \bibinfo{booktitle}{\emph{{WWW} '22: The {ACM} Web Conference 2022, Virtual
  Event, Lyon, France, April 25 - 29, 2022}}. \bibinfo{publisher}{{ACM}},
  \bibinfo{pages}{3376--3386}.
\newblock


\bibitem[Yuan et~al\mbox{.}(2014)]%
        {yuan14}
\bibfield{author}{\bibinfo{person}{Quan Yuan}, \bibinfo{person}{Gao Cong},
  {and} \bibinfo{person}{Chin-Yew Lin}.} \bibinfo{year}{2014}\natexlab{}.
\newblock \showarticletitle{COM: A Generative Model for Group Recommendation}.
  In \bibinfo{booktitle}{\emph{Proceedings of the 20th ACM SIGKDD International
  Conference on Knowledge Discovery and Data Mining}} (New York, New York, USA)
  \emph{(\bibinfo{series}{KDD '14})}. \bibinfo{publisher}{Association for
  Computing Machinery}, \bibinfo{address}{New York, NY, USA},
  \bibinfo{pages}{163–172}.
\newblock
\showISBNx{9781450329569}


\bibitem[Zhang et~al\mbox{.}(2017)]%
        {engagement@17}
\bibfield{author}{\bibinfo{person}{Mingli Zhang}, \bibinfo{person}{Lingyun
  Guo}, \bibinfo{person}{Mu Hu}, {and} \bibinfo{person}{Wenhua Liu}.}
  \bibinfo{year}{2017}\natexlab{}.
\newblock \showarticletitle{Influence of customer engagement with company
  social networks on stickiness: Mediating effect of customer value creation}.
\newblock \bibinfo{journal}{\emph{International Journal of Information
  Management}} \bibinfo{volume}{37}, \bibinfo{number}{3}
  (\bibinfo{year}{2017}), \bibinfo{pages}{229--240}.
\newblock
\showISSN{0268-4012}


\end{thebibliography}

\appendix

\end{document}